\documentclass[12pt]{article}   
\usepackage{epsfig}
\usepackage[utf8]{inputenc} 
\usepackage{url}
\usepackage{here}
\usepackage{amssymb}
\usepackage{graphicx}
\usepackage{comment}
\usepackage{verbatim}
\usepackage{fancyvrb}

\VerbatimFootnotes

\begin{document}


\title{Sceptical combination of experimental results\\ using {\tt JAGS/rjags} \\
  with 
  application to the K$^{\pm}$ mass determination}
\author{Giulio D'Agostini \\
Universit\`a ``La Sapienza'' and INFN, Roma, Italia \\
{\small (giulio.dagostini@roma1.infn.it,
 \url{http://www.roma1.infn.it/~dagos})}
}
\date{}
\maketitle
\thispagestyle{empty}

\begin{abstract}
The  question of how to combine experimental results
that `appear' to be in mutual disagreement, treated
in detail years ago in a previous paper, is revisited.
The first novelty of the present note
is the explicit use of graphical models, in order to make
the deterministic and probabilistic links between
the variables of interest more evident. 
Then, instead of
aiming for results in closed formulae,
the integrals of interest are evaluated
by {\em Markov Chain Monte Carlo} (MCMC) sampling, with
the algorithms (typically {\em Gibbs Sampler})
implemented in the package {\tt JAGS}
(``Just Another Gibbs Sampler'').
For convenience, the {\tt JAGS} functions are
called from {\tt R} scripts, thus gaining
the advantage given by the rich collection of mathematical, statistical
and graphical functions included in the {\tt R} installation.
The results of the previous paper are thus easily re-obtained
and the method is applied to the determination
of the charged kaon mass. This note, based on lectures
to PhD students and young researchers has been written
with a didactic touch, and the relevant {\tt JAGS/rjags}
code is provided. (A {\em curious bias} arising from the sequential
application of the $\sqrt{\chi^2/\nu}$ scaling prescription to
`apparently' discrepant results, found
here, will be discussed in more detail in a separate paper.)
\end{abstract}

\vspace{0.3cm}
{\small
\begin{flushright}
{\sl ``\ldots to emancipate us from the } \\
{\sl capricious ipse dixit of authority'' } \\
{(J.H. Newman)}\\
\end{flushright}
}
{\small
\begin{flushright}
{\sl ``Use enough common sense to know } \\
{\sl when ordinary common sense does not apply'' } \\
{(I.J. Good's {\it guiding principle of all science})}\\
\end{flushright}
}

\section{Introduction}
\begin{figure}[!t]
\centering\epsfig{file=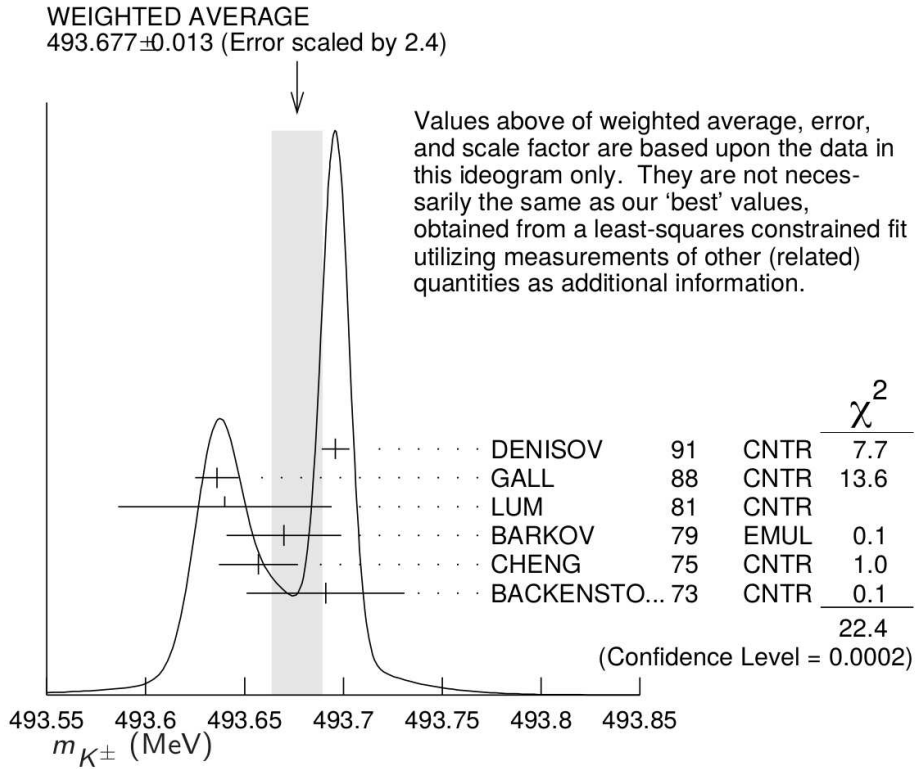,clip=,width=0.8\linewidth}
\caption{\small \sf Charged kaon mass from several experiments
  as summarized by the PDG~\cite{PDG2019}. Note that besides
  the `error' of 0.013\,MeV, obtained by a $\times 2.4$ scaling,
  also an `error' of  0.016\,MeV is provided,
  obtained by a $\times 2.8$ scaling. The two results are
  called `OUR AVERAGE' and 'OUR FIT', respectively.
}
\label{fig:PDF2019}
\end{figure}

It is not rare the case in which experimental
results `appear' to be in mutual disagreement. The quote marks
are mandatory, as a reminder that  
also very improbable events might by nature occur.\footnote{Remember
  that all events
  of our life {\em were} indeed VERY improbable, if observed with enough detail,
because they are just points in a high dimensional configuration space!}
The fact that they `appear' to us
in mutual disagreement is because we know by
experience that uncertainties\footnote{For the meaning of {\em error}
   and {\em uncertainty} see \cite{ISO} and \cite{ISOD}. Hereafter
   `error' in quote marks is to remind that the noun refers
   in reality to {\em uncertainty}, or, more precisely,
   {\em standard uncertainty}.}
might be underestimated,
systematic errors overlooked, theoretical corrections not (properly)
taken into account, 
 or even mistakes of
 different kinds having possibly
 been made in building/running the experiment or in the
 data handling.
It is enough to browse the PDG~\cite{PDG2019} to find
cases of this kind, as the one of Fig.~\ref{fig:PDF2019}
concerning the mass of the charged kaon, whose values,
as selected by the PDG,
are reported in Tab.~\ref{tab:masseK_PDG}.\footnote{Details
  can be found in the 2000 edition of the PDG~\cite{PDG2000}.
  Moreover, comparing the two editions of the PDG and taking
  into account that not always the details of the experiment are
  publicly available, it is clear that a {\em serious} work
  to determine at best the charged kaon mass goes beyond the {\em aim of this paper},
  being {\em mainly methodological}.
  Nevertheless, 
  the uncertainty reported for the 5th result of table
  Tab.~\ref{tab:masseK_PDG} is not a good account of the
  experimental result, 
  as it will be discussed later on in this paper.\label{fn:PDG2000}}
\begin{table}[t]
\begin{center}
  \begin{tabular}{|c|l|c|c|c|c|}
    \hline
 &   \multicolumn{1}{|c|}{Authors} & pub. year & central value $[d_i]$ & uncertainty $[s_i]$\\
 $i$            &        &   &    (MeV)      &    (MeV)    \\
    \hline
 $1$ &  G. Backenstoss et al. \cite{Kmass73}   & 1973 &  493.691 & 0.040 \\
 $2$ &     S.C. Cheng et al. \cite{Kmass75}        & 1975 & 493.657  & 0.020 \\
 $3$ &     L.M. Barkov et al.\cite{Kmass79}       & 1979 &  493.670  &  0.029 \\
 $4$ &     G.K. Lum et al. \cite{Kmass81}       & 1981   &  493.640  &   0.054 \\
 $5$ &     K.P. Gall et al. \cite{Kmass88}      &  1988  & 493.636   &    0.011 (*) \\
 $6$ &     A.S. Denisov et al. \cite{Kmass91}   &  1991  & 493.696  &  $[${\em 0.0059}$]$ \\
   &  \& {\em Yu.M. Ivanov} \cite{Kmass92}  & 1992  & $[${\em same}$]$&  0.007 \\
  \hline
\end{tabular}
  \caption{\small \sf Experimental values of the charged kaon mass,
    limited to those taken into account
  by the 2019 issue of PDG~\cite{PDG2019} (see footnote \ref{fn:PDG2000}
  for remarks).
}
\label{tab:masseK_PDG}
\end{center}
\end{table}

\begin{figure}[!t]
\centering\epsfig{file=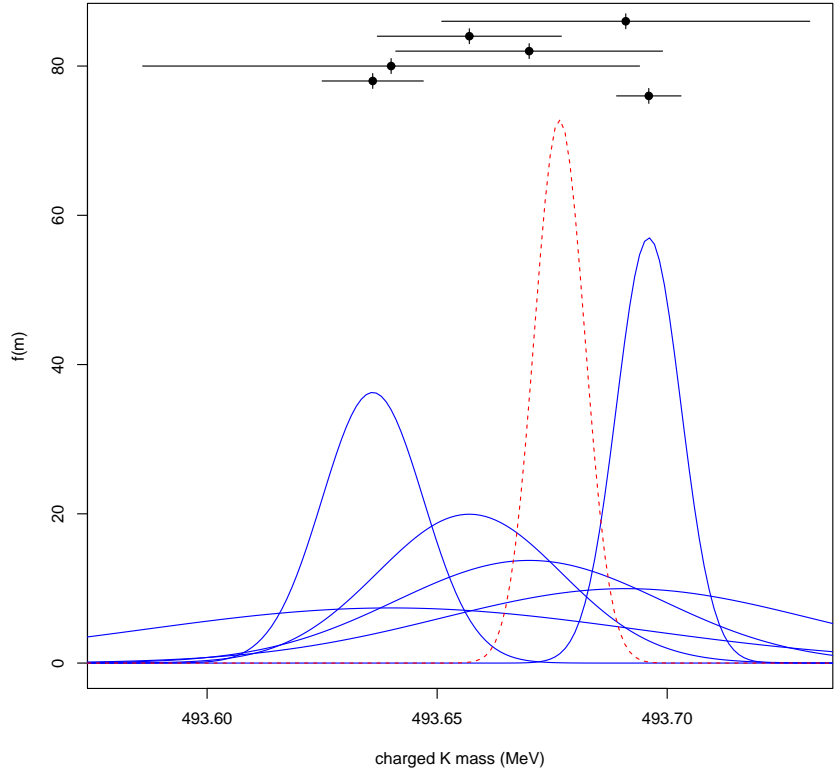,clip=,width=0.76\linewidth}
\caption{\small \sf Graphical representation of the results on the charged kaon mass
  of Tab.~\ref{tab:masseK_PDG} (solid lines). The dashed red Gaussian shows the result
  of the {\em naive} standard combination (see text).}
\label{fig:NaiveCombination}
\end{figure}

The usual probabilistic interpretation\footnote{Note that
  this interpretation is valid, under hypotheses which generally
  hold, especially if $s_i/d_i \ll 1$ (as it happens in this case),
  even if the results were produced with frequentistic methods
  that do not contemplate the possibility of attributing
  probabilities to the values of physics
  quantities. In fact, most results obtained using {\em standard statistics}
  ('frequentistic')  are based on the analysis of the so called
  {\em likelihood} around its maximum. And they can then be easily turned into
  probabilistic results
  (see e.g. \cite{BR}, in particular section 12.2.1
  and the related figure 12.1). }
of the results is that each experiment
provides a probability density function (pdf) centered in $d_i$
with standard deviation $s_i$, as shown by the solid lines 
of Fig.~\ref{fig:NaiveCombination}. 
The standard way to combine the individual results consists in 
calculating the weighted average, with weights equal to $1/s_i^2$, that is
\begin{eqnarray}
d_w &=& \frac{\sum_id_i/s_i^2}{\sum_i1/s_i^2} 
\label{eq:media}\\
s_w &=& \left(\sum_i1/s_i^2\right)^{-\frac{1}{2}}\,,
\label{eq:sigma}
\end{eqnarray}
which, applied to the values of Tab.~\ref{tab:masseK_PDG},
yields $d_w = 493.6766\,$MeV and  $s_w = 0.0055\,$MeV, i.e. 
a charged kaon mass of $493.6766\pm 0.0055\,$MeV,\footnote{In most cases
I stick here to two digits for the {\em standard uncertainty}.} graphically
shown in Fig.~\ref{fig:NaiveCombination} with a dashed red Gaussian.
The outcome `appears' suspicious because the probability mass is concentrated 
in the region less preferred by the individual more precise results,
as also emphasized in the {\em ideogram} of Fig.~\ref{fig:PDF2019},
on the meaning of which we shall return in section \ref{sec:JAGSresults}.

As a matter of fact, a situation of this kind is not impossible, but
nevertheless, there is a natural tendency to believe  that
there must be something not properly taken  into account by one or more
experiments. Told with a dictum attributed to a famous Italian politician,
{\em ``a pensar male degli altri si fa peccato ma spesso ci si
  indovina''}.\footnote{{\em ``To think badly would be to sin,
    but very often one gets it right''}$^{(*)}$.
Most Italians attribute it to Giulio Andreotti,
but it seems due no less then to a pope~\cite{AndreottiPioXI}.\\
$^{(*)}$\url{https://forum.wordreference.com/threads/a-pensare-male-si-fa-peccato-ma-spesso-ci-si-azzecca.2397506/}
}
\begin{figure}[!t]
\centering\epsfig{file=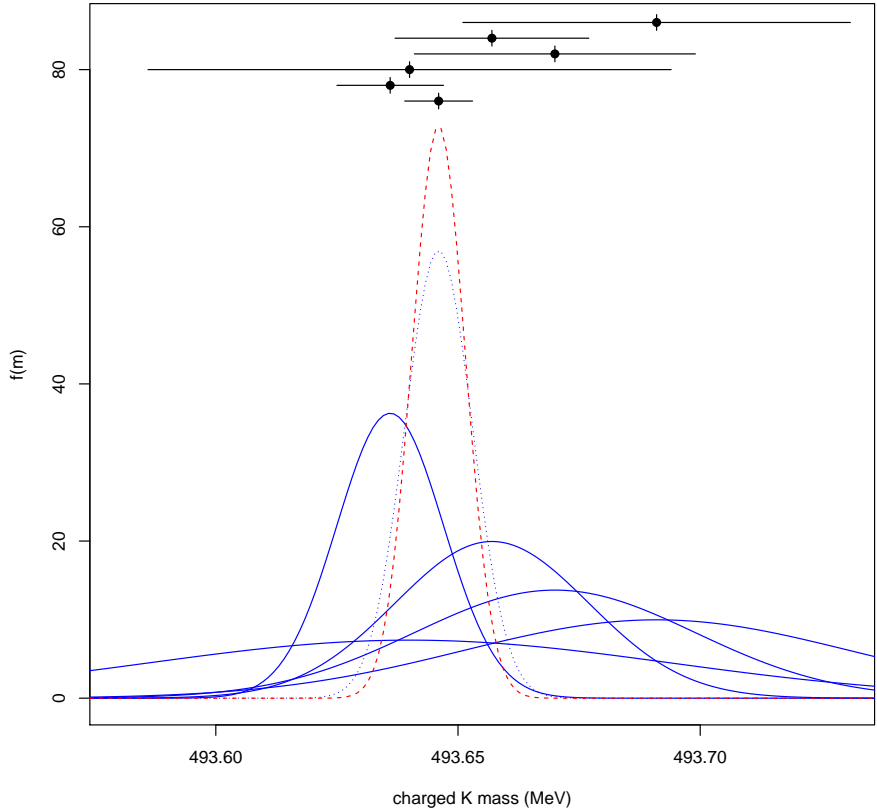,clip=,width=0.8\linewidth}
\caption{\small \sf Same as Fig.~\ref{fig:NaiveCombination} but with one
result {\em arbitrarily} shifted by $-50\,$keV (dotted line).}
\label{fig:NaiveCombinationAndreotti}
\end{figure}
For example, looking at Fig.~\ref{fig:NaiveCombination}, one is
strongly tempted to 
lower, {\em just as an exercise}, the highest value
by 50\,keV,\footnote{Value just decided by eye
  looking at the figure with some experienced colleagues, and not resulting
  from fits or optimizations of any kind.}
thus getting the excellent overall agreement shown in
Fig.~\ref{fig:NaiveCombinationAndreotti}
(shifted Gaussian plotted with a dotted gray line), 
yielding a combined mass value of $493.6460 \pm 0.0055\,$MeV.
And the question would be settled. But this sounds at least
unfair.
  In particular because we are aware, from the history of measurements,
  of a kind of `inertia' of new results to different from old
  ones
  -- but sometimes the new results moved `too far' from the
  old ones and the presently {\em accepted} value
  lies somewhere in the middle.
\begin{figure}[!t]
\centering\epsfig{file=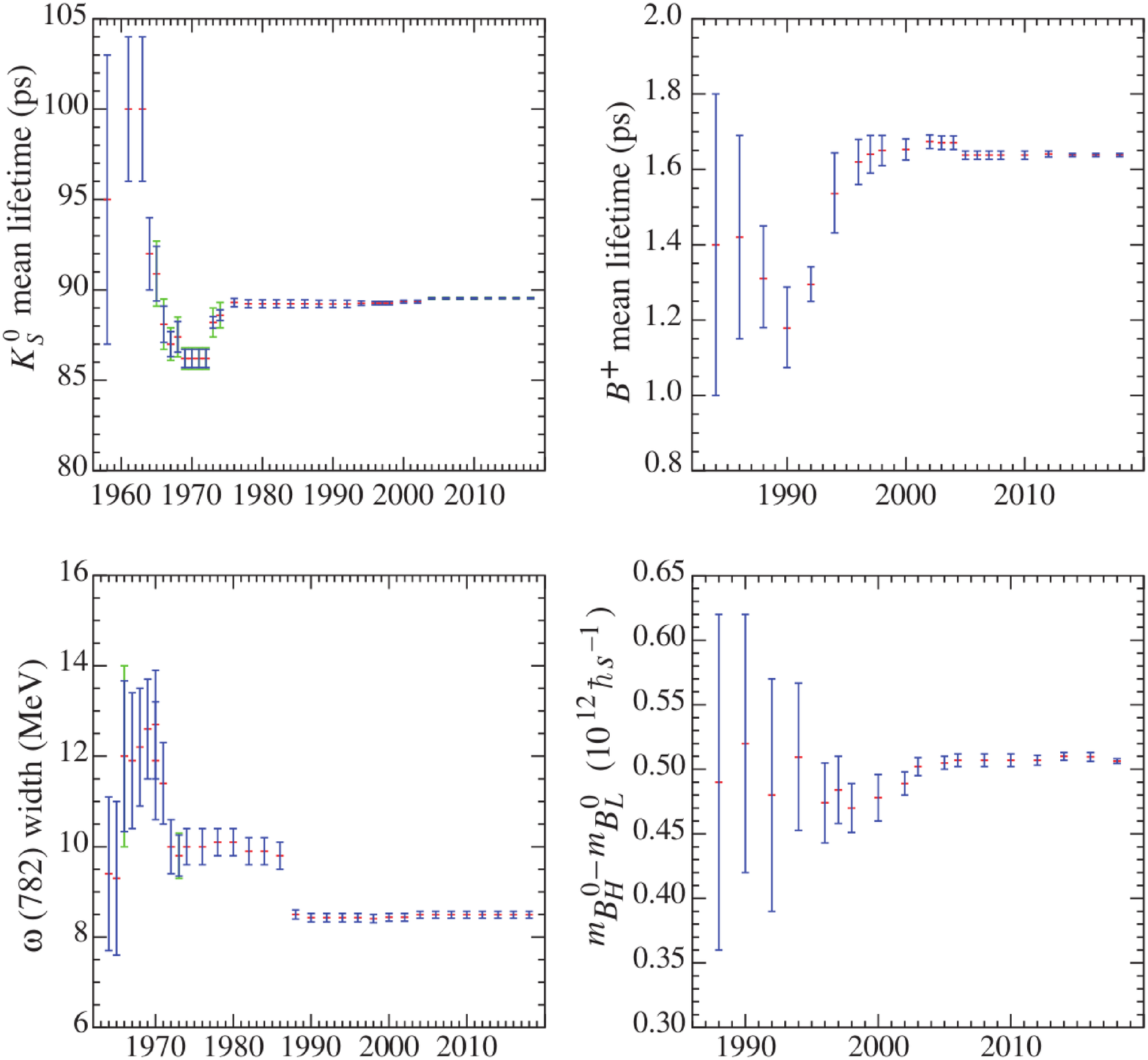,clip=,width=0.87\linewidth}
\caption{\small \sf Some history plots from the
  PDG~\cite{PDG2019,PDG_HistoryPlots}.}
\label{fig:HistoryPlots}
\end{figure}
Figure \ref{fig:HistoryPlots} shows some of the history plots
traditionally reported by the PDG~\cite{PDG_HistoryPlots}.

  In such a state of uncertainty, probability theory can help us
  in building up a model in which the values about which we are in doubt
  are allowed to vary from the nominal ones. Obviously, the model is not
  unique, as not unique are the probability distributions that can be used.
  Following then \cite{sceptical1999}, inspired to \cite{Dose}, these
  are the criteria followed and some (hopefully shareable)
  {\em desiderata}:
  \begin{itemize}
  \item the quantities on which we focus our doubts are
    the reported uncertainties $s_i$, assumed to have the meaning of
    {\em standard uncertainty}~\cite{ISO,ISOD}, as commented
    above;\footnote{A possible alternative would be to allow a shift
      of the measured quantity. However this model seems unable to
      yield multimodal final pdf's, which is, in my opinion, one of the
      {\em desiderata} of the model, as stated here. 
      Perhaps the question requires further study
      but, given the limited aims of this paper, I prefer to stick
    for the moment to the models of Refs.~\cite{sceptical1999,Dose}.}
  \item the `true' standard deviations $\sigma_i$ are related
    to $s_i$ by a factor $r_i$ (one for each experiment), i.e.
    \begin{eqnarray*}
      \sigma_i &=& r_i\,s_i \\
      d_i      &\sim & {\cal N}(\mu, \sigma_i)\,,
    \end{eqnarray*}
    where the last notation means that $d_i$ is described
    by a Gaussian (`normal') distribution centered on the `true'
    value of the quantity of interest, generically indicated by $\mu$,
    with standard deviation $\sigma_i$;
  \item all experiments are treated {\em democratically} and {\em
    fairly}, i.e. our {\em prior belief} of each $r_i$ 
    has  expected value equal to 1, and its prior distribution
    does not depend on the experiment:
    \begin{eqnarray*}
      f(r_1\,|\,I) &=& f(r_2\,|\,I)\  = \ f(r_3\,|\,I)\  =\  \cdots \\
      \mbox{E}[r_i\,|\,I] &=& 1\,,
    \end{eqnarray*}
    where `$I$' stands for the background {\em status of information}
    (probability is \underline{always} conditional probability!);
  \item but we are {\em sceptical}, and hence each $r_i$ has {\em à priori}
    a wide range of possibilities
    described by a suitable (easy to handle) probability distribution,
    the details of which will be give later -- we just anticipate that
    we take a prior 100\% standard uncertainty on $r_i$,
    i.e. $\sigma(r_i\,|\,I)/\mbox{E}[r_i\,|\,I]=1$;
  \item one of the  desiderata of the model is that the {\em posterior pdf}
    of the physical quantity of interest
    {\em should not be limited to a Gaussian} and could even be multimodal
    if the individual results cluster in different regions;
    or it could be narrower than the pdf obtained by the
    standard weighted average, if the individual results tend to overlap
    `too much'
    (see e.g. Figs.~4 and 5 of
    Ref.~\cite{sceptical1999});
  \item finally, once the parameters of  $f(r_i\,|\,I)$
    are defined on bench marks and checked against `reasonable'
    variation (as done in Figs.~4 and 5 of
    Ref.~\cite{sceptical1999}), {\em fine tuning and cherry peaking
    of the individual results to be included in the
    combination should be avoided} (unless we have good reasons
    to mistrust some results).
  \end{itemize}
  Once the model has been built, we can easily write down
  the multidimensional
  probability pdf
  $f(\underline{d},\mu,\underline{r}\,|\,\underline{s},I)$,
  of all the variables of interest
  (the `observed' $d_i$ and 
  the uncertain values $\mu$ and $r_i$'s \,-- the $s_i$ will
  be instead considered as fixed conditions, as it will be clear in a while;
   $\underline{d}$ stands for all the $d_i$, and so on). 
    
  Once the multi-dimensional pdf has been
  settled, writing down the {\em unnormalized}
  pdf of the uncertain quantities,
  \begin{eqnarray*}
    \widetilde{f}(\mu,\underline{r}\,|\,\underline{d},\underline{s},I)
    &\propto& f(\mu,\underline{r}\,|\,\underline{d},\underline{s},I)\,,
  \end{eqnarray*}
  is straightforward, as we shall see in a while. But, differently from
  \cite{sceptical1999}, the rest of the technical work (normalization,
  marginalization and calculation of the moments of interest) will
  be done here by sampling, i.e. by Monte Carlo, and the use of a
  suitable software package will make the task rather easy.

  But, before we build up the model of interest, let us start
  with a simpler one, in which 
  we fully trust the reported standard uncertainty, i.e.
  we assume  $f(r_i\,|\,I) = \delta(1)$, and hence $\sigma_i=s_i$.
  We also take for the prior,
  following Gauss\,\cite{Gauss,Gauss_trasl},
  a flat distribution of $\mu$
  in the region of interest.\footnote{For the Gauss' use of what we would
    nowadays call a {\em Bayesian reasoning}, starting form
    the concept of probabilities of the true value, see
    Section 6.12 of Ref.~\cite{BR} based
    on  Section III of Book II of Ref.~\cite{Gauss}
    $[$see Ref.~\cite{GdA_history} for 
      details  on the missing steps between Eq.(6.53) and Eq.(6.54)$]$.
    Here I just want to comment on the meaning of a `flat' prior, which does
    not imply
    that it has to be interpreted as strictly constant all over the real axis.
    With this respect it is interesting the comment that Gauss
    adds after he derived the `Gaussian' as the error function
    characterized by {\em good mathematical behavior} and such that
    the {\em posterior gets its maximum in correspondence
      of the arithmetic average},
    in the case of independent measurements characterized by the same
    error probability distribution:
      {\sl ``The function just found $[${\rm the `Gaussian'}$]$ cannot,
        it is true, express rigorously  the probabilities of the errors: for
        since the possible errors are in all cases confined
        within certain limits, the probability of errors exceeding those
        limits ought always be zero, while our formula always gives some value.
        However, this defect, which every analytical function must, from its nature,
        labor under, is of no importance in practice, because the value of
        function decreases so rapidly, when $h\,\Delta$
        $[${\rm `$(x_i-\mu)/\sigma$', in modern notation}$]$
        has acquired a considerable magnitude, that it can safely be considered
        as vanishing. Besides, the nature of the subject never admits of assigning with
        absolute rigor the limits of error.''}~\cite{Gauss_trasl}
    \label{fn:Gauss}
  }
  
\section{Standard combination from a probabilistic perspective}
\begin{figure}[!t]
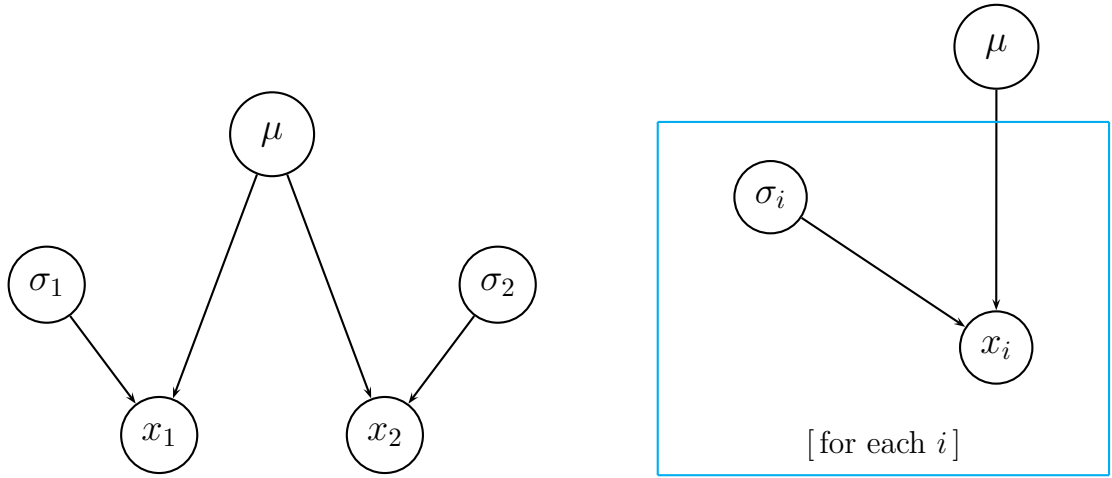

\begin{center}
\begin{tabular}{cc}    
  \epsfig{file=standcomb2.epsi,clip=,}\hspace{0.5cm} &
  \hspace{0.5cm} \epsfig{file=standcomb.epsi,clip=,}
\end{tabular}  
\end{center}  
\caption{\small \sf Graphical model behind the standard combination.}
\label{fig:StandardComb}
\end{figure}

Let us start with just two experimental outcomes,\footnote{In this
  introductory section we use  $x_i$  to indicate an individual
  observation, while in general the $d_i$ of Tab.~\ref{tab:masseK_PDG}
  are results of `statistical analyses' based
  on many direct `observations'.}
    $x_1$ and $x_2$, resulting from the
uncertain `true' value $\mu$ (what we are interested in)
when measured in two independent experiments having Gaussian
error functions with standard deviations $\sigma_1$ and $\sigma_2$,
as sketched in the left hand graph of Fig.~\ref{fig:StandardComb}.
That is
\begin{eqnarray*}
  x_1      &\sim & {\cal N}(\mu, \sigma_1)\\
  x_2      &\sim & {\cal N}(\mu, \sigma_2)\,.  
\end{eqnarray*}
The general case, with many measurements of the same $\mu$,
is shown on the right hand graph of the same figure. 
From a probabilistic point of view our aim will be to
assess, with a probability distribution, the intervals
where we believe $\mu$ lies with different probabilities, 
that is to arrive to
\begin{eqnarray*}
  f(\mu\,|\,x_1,x_2,\ldots,\sigma_1,\sigma_2,\ldots,I) \equiv
  f(\mu\,|\,\underline{x},\underline{\sigma}\ldots,I).
\end{eqnarray*}  
The pdf of interest can be ({\em in principle}) obtained easily
if we knew the joint pdf of all the quantities of interest,
that is $f(\mu,\underline{x}\,|\,\underline{\sigma},I)$. In fact,
we just need to apply
a well know general rule of probability theory
(remember that in this first example
the $\sigma_i$ are just fixed conditions, although in general 
they might become subject to inference too, as we shall see later):
\begin{eqnarray*}  
  f(\mu\,|\,\underline{x},\underline{\sigma},I)  &=&
  \frac{f(\mu,\underline{x}\,|\,\underline{\sigma},I)}
       {f(\underline{x}\,|\,\underline{\sigma},I)}\,.
\end{eqnarray*}  
At this point the reader might be scared by two reasons:
the first is how to build up the joint pdf
$f(\mu,\underline{x}\,|\,\underline{\sigma},I)$;
the second is how to perform the integral over $\mu$ (`{\em marginalization}')
in order to get the denominator.\footnote{Let us remind that, in general,
  $f(\underline{x}\,|\,\underline{\sigma},I) =
  \int_{-\infty}^{+\infty}f(\mu,\underline{x}\,|\,\underline{\sigma},I)\,d\mu$\,.}

The first good news is that, given the model
(those of Fig.~\ref{fig:StandardComb} or the more complicate ones
we shall see later), {\em the denominator is just a number},
in general difficult to calculate, but  \underline{just a number}. This means
that we can rewrite the previous equation as
\begin{eqnarray}  
  f(\mu\,|\,\underline{x},\underline{\sigma},I)  &\propto&
  f(\mu,\underline{x}\,|\,\underline{\sigma},I)\,.
  \label{eq:nn_f_mu}
\end{eqnarray}  
As next step, we can follow two strategies:
\begin{itemize}
\item calculate the normalization factor at the end, either
  analytically or numerically;
\item sample the unnormalized distribution by Monte Carlo
  techniques, in order to get the shape of
  $f(\mu\,|\,\underline{x},\underline{\sigma},I)$ and to calculate
  all moments of interest.
\end{itemize}
The second good news is that the multidimensional joint pdf
can be easily written down using the well known
probability theory theorem known as
{\em chain rule}. Indeed, sticking to the model with just two
variables, we can apply the chain rule in different ways.
For example,
beginning from the most pedantic one, we have
\begin{eqnarray*}  
  f(\mu,x_1,x_2\,|\,\sigma_1,\sigma_2,I) 
   &=&  f(\mu\,|\,x_1,x_2,\sigma_1,\sigma_2,I)\cdot
       f(x_1\,|\,x_2,\sigma_1,\sigma_2,I) \cdot
       f(x_2\,|\,\sigma_1,\sigma_2,I)\,.
\end{eqnarray*}
But this writing does not help us,
since it requires 
$ f(\mu\,|\,x_1,x_2,\sigma_1,\sigma_2,I)$, which it is
precisely what we aim for.
It is indeed much better, with an eye to
Fig.~\ref{fig:StandardComb}, a {\em bottom up} approach (in the following
equation the order of the arguments in the left side term
has been changed to make  the correspondence
between the two writings easier), that is
\begin{eqnarray*}  
  f(x_1,x_2,\mu,\,|\,\sigma_1,\sigma_2I)  & = &
  f(x_1\,|\,x_2,\mu,\sigma_1,\sigma_2,I) \cdot
  f(x_2\,|\,\mu,\sigma_1,\sigma_2,I) \cdot
  f(\mu\,|\,\sigma_1,\sigma_2,I)\,. 
\end{eqnarray*}
This equation can be
further simplified if we note that each $x_i$ depends
directly only on $\mu$ and $\sigma_i$, while
$\mu$ does not depend (at least in usual measurements)
on $\sigma_1$ and $\sigma_2$. We get then
\begin{eqnarray}  
  f(x_1,x_2,\mu,\,|\,\sigma_1,\sigma_2,I)  & = &
  f(x_1\,|\,\mu,\sigma_1,I) \cdot
  f(x_2\,|\,\mu,\sigma_2,I) \cdot
  f(\mu\,|\,I)\,.
  \label{eq:chain_rule_x1_x2_mu}
\end{eqnarray}
We can easily generalize this equation, in the case of many observations
described in the right hand graph of Fig.~\ref{fig:StandardComb},
rewriting it as
\begin{eqnarray}  
  f(\underline{x},\mu,\,|\,\underline{\sigma})  & = &
  \left[ \prod_i f(x_i\,|\,\mu,\sigma_i)\right]\cdot f_0(\mu)
  \label{eq:fxy_1} 
\end{eqnarray}
where the index $i$ runs through all the observations
and the symbol `$I$' indicating the background state of information 
has been dropped, using `$f_0(\mu)$' 
for the {\em initial distribution} (`prior') of $\mu$.
Finally, taking (for the moment)
for $f_0(\mu)$ a
{\em practically flat} distribution in the region of interest
(see footnote \ref{fn:Gauss}), making use of
Eq.~(\ref{eq:nn_f_mu}) and of the symbol $f_{\cal N}$
to indicate {\em normal} (i.e. Gaussian) error functions, 
we get
\begin{eqnarray}  
  f(\mu\,|\,\underline{x},\underline{\sigma},f_0(\mu)=k)  &\propto&
  f(\underline{x},\mu,\,|\,\underline{\sigma}) \ \propto \
   \prod_i f_{{\cal N}}(x_i\,|\,\mu,\sigma_i)\,.
   \label{eq:nnG_f_mu}
\end{eqnarray}
Using the explicit expression of the Gaussian and neglecting
all multiplicative factors that do not depend on $\mu$,
we get
\newpage
\begin{eqnarray}  
  f(\mu\,|\,\underline{x},\underline{\sigma},f_0(\mu)=k)   &\propto&  \prod_i \exp\left[-\frac{(x_i-\mu)^2}{2\,\sigma_i^2}\right]
 \label{eq:nnG_f_mu_Norm}   \\
   &\propto&   \exp\left[-\sum_i\,\frac{(x_i-\mu)^2}{2\,\sigma_i^2}\right] \nonumber \\
   &\propto&   \exp\left[-\frac{1}{2}
     \sum_i\,\frac{x_i^2-2 x_i \mu + \mu^2}{\sigma_i^2}
     \right]  \nonumber \\
    &\propto&   \exp\left[-\frac{1}{2}
     \sum_i\,\left(\frac{x_i^2}{\sigma_i^2}-2\, \frac{x_i}{\sigma_i^2}\, \mu +
     \frac{\mu^2}{\sigma_i^2}\right)
     \right] \nonumber \\
   &\propto&   \exp\left[-\frac{1}{2}\cdot
     \frac{\sum_i 1/\sigma_i^2}{\sum_i 1/\sigma_i^2} \cdot 
     \left(\sum_i\frac{x_i^2}{\sigma_i^2}-2\, \left(\sum_i\frac{x_i}{\sigma_i^2}\right)\, \mu +
     \left(\sum_i\frac{1}{\sigma_i^2}\right)  \mu^2 \right)
     \right] \nonumber \\
 &\propto&  \exp\left[-\frac{1}{2}\cdot \left(\sum_i 1/\sigma_i^2\right)
   \cdot\left( \overline{x^2} - 2\,\overline{x}\,\mu
     + \mu^2  \right)
     \right]  \nonumber \\
    &\propto&  \exp\left[- \frac{ \overline{x^2} - 2\,\overline{x}\,\mu
     + \mu^2}{2/(\sum_i 1/\sigma_i^2)}
     \right]  \label{eq:lik_con_x2ave}\\
    &\propto&  \exp\left[- \frac{ - 2\,\overline{x}\,\mu
     + \mu^2}{2\,\sigma_C^2}
     \right] \,, \label{eq:lik_no_x2ave}
\end{eqnarray} 
where
\begin{eqnarray}
  \overline{x} &=& \frac{\sum_i\,x_i/\sigma_i^2} {\sum_i 1/\sigma_i^2}
  \label{eq:media_pesata}\\
  && \nonumber \\
  \overline{x^2}& =& \frac{\sum_i\,x_i^2/\sigma_i^2} {\sum_i 1/\sigma_i^2}\nonumber\\
  && \nonumber \\
  \sigma_C^2 &=& \frac{1}{\sum_i 1/\sigma_i^2}\,.
  \label{eq:sigma_media}  
\end{eqnarray}
Note that the mean of the squares 
$\overline{x^2}$ has been taken out of the exponent
$[$\,step from Eq.~(\ref{eq:lik_con_x2ave}) to Eq.~(\ref{eq:lik_no_x2ave})\,$]$
because
$\exp[-\overline{x^2}/(2\sigma_C^2)]$ does not depend on $\mu$
and therefore it can be absorbed in the normalization constant.
For the same reason we can multiply Eq.~(\ref{eq:lik_no_x2ave}) by
$\exp[-\overline{x}^2/(2\sigma_C^2)]$, thus getting,
by {\em complementing the exponential},
\begin{eqnarray*}
  f(\mu\,|\,\underline{x},\underline{\sigma},f_0(\mu)=k)  &\propto&
    \exp\left[- \frac{ \overline{x}^2 - 2\,\overline{x}\,\mu
     + \mu^2}{2\,\sigma_C^2}
      \right] \nonumber \\
    && \nonumber \\
    &\propto&    \exp\left[- \frac { (\mu-\overline{x})^2}
                                   {2\,\sigma_C^2}
      \right] 
\end{eqnarray*}
We can now recognize in it, at first sight, 
a Gaussian distribution of the variable $\mu$ around $\overline{x}$,
with standard deviation $\sigma_C$, i.e.
\begin{eqnarray}
  f(\mu\,|\,\underline{x},\underline{\sigma},f_0(\mu)=k) &=&
  \frac{1}{\sqrt{2\,\pi}\,\sigma_C}\,  \exp\left[- \frac { (\mu-\overline{x})^2}
    {2\,\sigma_C^2} \right]
  \label{eq:G_f_mu_Norm}
\end{eqnarray}
with expected value  $\overline{x}$ and standard deviation $\sigma_C$.

Someone might be worried about the dependence of the inference on
the flat prior of $\mu$, written explicitly in Eq.~(\ref{eq:G_f_mu_Norm}),
but what really matters is that $f_0(\mu)$ does not vary much 
in the region of a few $\sigma_C$'s around $\overline{x}$.
Since in the software package that we are going to use, starting
from next section, an explicit prior is required,
let us try to understand
the {\em influence of a vague but not flat prior} in the resulting inference.
Let us model $f_0(\mu)$ with a Gaussian distribution having a
rather large $\sigma_0$ (e.g. $\sigma_0 \gg \sigma_i$)
and centered in $x_0 \approx {\cal O}(x_i)$. The result is that
 Eq.~(\ref{eq:nnG_f_mu_Norm}) becomes 
\begin{eqnarray}  
  f(\mu\,|\,\underline{x},\underline{\sigma})  &\propto&
  \prod_i \exp\left[-\frac{(x_i-\mu)^2}{2\,\sigma_i^2}\right]\cdot
  \exp\left[-\frac{(\mu-x_0)^2}{2\,\sigma_0^2}\right]\,.
 \label{eq:nnG_f_mu_Norm_f0}  
\end{eqnarray}
This is equivalent to add the extra term 
 $x_0$ with standard uncertainty $\sigma_0$, which
has then to be included in the calculation of $\overline{x}$
and $\sigma_C$ $[$technically the index $i$ in the sums  in
 Eqs.~(\ref{eq:media_pesata}) and (\ref{eq:sigma_media})
 run from 0 to $n$,
instead than from $1$ to $n$, being $n$ the number of measurements$]$.
But if $\sigma_0 \gg \sigma_i$ (more precisely
$ \frac{1}{\sigma_0^2} \ll \sum_{i=1}^n\frac{1}{\sigma_i^2}$)
and $x_0$ is `reasonable',
then
the extra contribution is irrelevant.

\section{Probabilistic combination achieved by Monte\\ Carlo sampling
  using {\tt JAGS} and {\tt rjags}}
\begin{figure}[!t]
\begin{center}
\begin{tabular}{c}    
  \epsfig{file=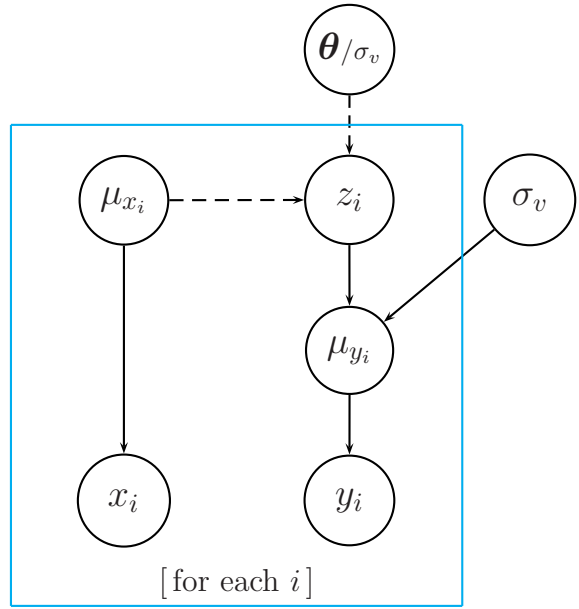,clip=,}
\end{tabular}  
\end{center}  
\caption{\small \sf Graphical model for a non trivial fit with errors
on both axes~\cite{fits}.}
\label{fig:fits}
\end{figure}

The case just analyzed is so simple that, even without getting
the solution in closed form, it is enough to plot the unnormalized
pdf (\ref{eq:nnG_f_mu_Norm}) to understand what is going
on and to get `somehow' mean value and standard deviation.
The problem becomes more serious in the case we want to make
a multidimensional inference, taking into account also the
correlations between the quantities of interest, as
for example in the fit model of Fig.~\ref{fig:fits},
taken from Ref.~\cite{fits}.

Nowadays
the most general
way to handle problems of this kind is by sampling the unnormalized
posterior distribution by a Markov Chain Monte Carlo (MCMC),
using a suitable algorithm
(see Ref.~\cite{Andrieu} for an introduction -- given the
imaginable interest of the subject in many fields,
much more can be found searching on the web;
in particular, particle physicists might be interested
in {\tt BAT}~\cite{BAT}, the
{\em Bayesian Analysis Toolkit}). 
%
Perhaps (said by a non expert)
the most powerful MCMC algorithm is the so called {\em Metropolis}
(with variants),
but for the kind of problem in which we are interested in this paper
the most convenient one is the so called
{\em Gibbs Sampler},\footnote{Talking
  about the Gibbs sampler algorithm applied in
  probabilistic inference (and forecasting) it is impossible 
not to to mention the {\em BUGS project} ~\cite{BUGSpaper}, the acronym staying
for \underline{B}ayesian inference \underline{u}sing
\underline{G}ibbs \underline{S}ampler, that has
been a kind of revolution in Bayesian analysis,
decades ago limited to simple cases because of computational problems
(see also Section 1 of \cite{JAGS}).
In the project web site~\cite{BUGSsite}
  it is possible to find packages with excellent Graphical User Interface,
  tutorials and many examples \cite{BUGSexamples}, which, although
  far from the typical interests of physicists,
  might help to understand the underlying reasoning
  and the model language, practically the same used by {\tt JAGS}.
}
although
it has some limitations on the conditional distributions it
can handle
(see \cite{Andrieu} and \cite{BUGSsite} for details).

Instead of writing our own code, which would be
anyway rather easy for
our simple problem, we are going to use the program
{\tt JAGS}~\cite{JAGS},
born as an open source, multi-platform clone of {\tt BUGS}.
{\tt JAGS} does not come with a graphical interface, so it is convenient
to use it within a more general framework like {\tt R}~\cite{R}
via the package {\tt rjags}~\cite{rjags} (those who are familiar with
Python might want to use {\tt pyjags}~\cite{pyjags}).

\subsection{Gibbs sampling in the spaces $(\mu,\sigma)$ and $(\mu,\tau)$}
Before we write down the model to solve our little problem
described by the {\em graphical model} of Fig.~\ref{fig:StandardComb},
some words on the Gibbs sampling are needed, also to understand
why this kind
of programs do not use $\sigma$ as second parameter of the Gaussian,
but rather $1/\sigma^2$, traditionally indicated by $\tau$.

We have seen that if we have a problem with Gaussian error functions
and a flat prior on $\mu$, then 
the posterior of $\mu$ is still a Gaussian, and it remains Gaussian
also if we assign to $\mu$ a Gaussian prior characterized by $x_0$
and $\sigma_0$, a flat prior being recovered
for $\sigma_o\rightarrow\infty$ (and allow me to draw again your
attention on footnote~\ref{fn:Gauss}).
Let us see what happens if we are also in condition
of uncertainty concerning $\sigma$,
assumed to be the same in all $n$ measurements (typical problem
of when we collect a sample a measurements under apparently
the same conditions and we are interested in inferring
both $\mu$ and $\sigma$). The graphical model
is still the one on the right hand side of  Fig.~\ref{fig:StandardComb},
but with $\sigma_i=\sigma$ for all $i$.
The analogue of Eq.~(\ref{eq:nn_f_mu}) is now
\begin{eqnarray}  
  f(\mu,\sigma\,|\,\underline{x},I)  &\propto&
  f(\mu,\sigma,\underline{x}\,|\,I)\,.
  \label{eq:nn_f_mu_sigma}
\end{eqnarray}  
Applying the chain rule only to $x_1$ and $x_2$,
to begin,
and noting that $\mu$ and $\sigma$ do not depend on each other,
as it is usually the case,\footnote{But in frontier research it is not difficult
  to imagine cases in which this is not true.}
we have, instead of Eq.~(\ref{eq:chain_rule_x1_x2_mu}),
 \begin{eqnarray}  
  f(x_1,x_2,\mu,\sigma\,|\,I)  & = &
  f(x_1\,|\,\mu,\sigma,I) \cdot
  f(x_2\,|\,\mu,\sigma,I) \cdot
  f(\mu\,|\,I)\cdot f(\sigma\,|\,I) \,.
  \label{eq:chain_rule_x1_x2_mu_sigma}
\end{eqnarray}  
Equations
(\ref{eq:fxy_1}) becomes then, also extending Eq.~(\ref{eq:chain_rule_x1_x2_mu_sigma})
to all $x_i$,
\begin{eqnarray*}  
  f(\underline{x},\mu,\sigma\,|\,I)  & = &
  \left[ \prod_i f(x_i\,|\,\mu,\sigma,I)\right]\cdot f(\mu\,|\,I)
  \cdot f(\sigma\,|\,I)  \nonumber
  \\
  &=&  \left[ \prod_i f_{{\cal N}}(x_i\,|\,\mu,\sigma)\right]\cdot
  f_0(\mu)\cdot f_0(\sigma) \,. \nonumber 
\end{eqnarray*}
Now, if {\em for some reasons} we fix $\sigma$ to the
hypothetical value  $\sigma^*$ (and for simplicity we use a flat
prior for $\mu$) then we recover, without any calculation,
something similar to Eq.~(\ref{eq:G_f_mu_Norm}): 
\begin{eqnarray}
  f(\mu\,|\,\underline{x},\sigma^*,f_0(\mu)=k) &=&
  \frac{1}{\sqrt{2\,\pi}\,\sigma_C}\,  \exp\left[- \frac { (\mu-\overline{x})^2}
    {2\,\sigma_C^2} \right]
  \label{eq:G_f_mu_Norm_sigma*}
\end{eqnarray}
where now $\overline{x}$ is simply the arithmetic average
and  $\sigma_C = \sigma^*/\sqrt{n}$. If we had taken into account
a prior $f_0(\mu)$
modeled by a Gaussian, then $f(\mu\,|\,\underline{x},\sigma^*)$
would still be a Gaussian, as we have seen above.
In particular, {\em it easy to sample by
  Monte Carlo} the
`random' variable $\mu$ described by Eq.~(\ref{eq:G_f_mu_Norm_sigma*}),
because it is rather easy to write a
Gaussian `random' number generator, or to use one of those available
in the mathematical libraries of programming languages. 

Let us now do the opposite exercise, the utility
of which will be clear in a while: imagine we are interested
in  $f(\sigma\,|\,\underline{x},\mu^*,f_0(\sigma)=k)$,
having imposed the condition $\mu=\mu^*$.
Equations (\ref{eq:nnG_f_mu}) and
(\ref{eq:nnG_f_mu_Norm}) are now turned into
(note the factor $1/\sigma$ in front of each exponent,
since it cannot be any longer absorbed in the normalization constant!)
\begin{eqnarray}  
  f(\sigma\,|\,\underline{x},\mu^*,f_0(\sigma)=k)  &\propto&
  f(\underline{x},\mu^*,\sigma) \ \propto \
   \prod_i f_{{\cal N}}(x_i\,|\,\mu^*,\sigma)
   \label{eq:nnG_f_sigma} \\
   &\propto&  \prod_i\,
   \frac{1}{\mathbf{\sigma}}\,
   \exp\left[-\frac{(x_i-\mu^*)^2}{2\,\sigma^2}\right] 
   \label{eq:nnG_f_sigma_Norm}   \\
   &\propto& \frac{1}{\mathbf{\sigma}^n}\,
   \exp\left[-\frac{\sum_i(x_i-\mu^*)^2}{2\,\sigma^2}\right]  \nonumber\\
   &\propto& \frac{1}{\mathbf{\sigma}^n}\,
   \exp\left[-\frac{K^2(\underline{x},\mu^*)}{\sigma^2}\right] \nonumber\,,
\end{eqnarray}
where $K^2(\underline{x},\mu^*)$ is a constant, given $\underline{x}$
and $\mu^*$, written in a way to remind that it is
by definition non negative. Unfortunately, opposite to the case
of $f(\mu\,|\,\overline{x},\sigma^*)$, this is an unusual form
in probability theory. But a simple change of variable
rescues us. In fact, if instead of $\sigma$ we use $\tau=1/\sigma^2$,
then the last equation becomes
\begin{eqnarray}  
  f(\tau\,|\,\underline{x},\mu^*,f_0(\sigma)=k)  &\propto&
  \tau^{\frac{n}{2}}\,\exp\left[-K^2(\underline{x},\mu^*)\cdot\tau\right]
   \nonumber
\end{eqnarray}
in which probability and statistics experts recognize
a Gamma distribution, usually written for the generic variable $z$ as
\begin{eqnarray}
  f(z\,|\,\alpha,\beta) &=&  \frac{\beta^{\,\alpha}}{\Gamma(\alpha)}\,
  z^{\alpha -1}\,e^{-\beta\,z}\,, \nonumber
\end{eqnarray}
where $\Gamma()$ is the Gamma function
(and hence the name of the distribution).
Therefore
\begin{eqnarray}  
  f(\tau\,|\,\underline{x},\mu^*,f_0(\sigma)=k)  &=&
  \frac{\beta^{\,\alpha}}{\Gamma(\alpha)}
  \,\tau^{\alpha-1}\,e^{-\beta\,\tau}
  \label{eq:Gamma_tau}
\end{eqnarray}
with $\alpha = 1 + n/2$ and
$\beta = K^2(\underline{x},\mu^*) = \sum_i(x_i-\mu^*)^2/2$.
Being this a well known probability distribution,
there are formulae available for the summaries
of interest.\footnote{My preferred {\em vademecum} of Probability
Distributions is the homonymous {\em app}~\cite{ProbabilityDistributions}.}
For example, expected value and variance
are given by $\alpha/\beta$ and $\alpha/\beta^2$, respectively.
But, moreover, there are Gamma random generators available, which
is what we need for sampling.

We are then finally ready to describe the {\em Gibbs sampler} algorithm,
applied to our two-dimensional case
(but it can be applied in higher dimensionality problems too):
\begin{itemize}
\item start choosing an {\em arbitrary} initial point  $(\mu_0,\,\tau_0)$
  in the $(\mu,\,\tau)$ plane;
\item extract `at random' a new value of $\mu$, let it be $\mu_1$,
  given $\tau_0$, from $f(\mu\,|\,\underline{x}, \tau_0)$;
\item extract then `at random'  a new value of $\tau$, let it be $\tau_1$,
  given $\mu_1$, from $f(\tau\,|\,\underline{x}, \mu_1)$;
\item extract then `at random'   a new value of $\mu$, let it be $\mu_2$,
   from $f(\mu\,|\,\underline{x}, \tau_1)$;
\item continue on, through the steps  $\tau_1 \rightarrow \mu_2
  \rightarrow \tau_2 \rightarrow \mu_3 \rightarrow
  \tau_3 \rightarrow \cdots\cdots
  \rightarrow \mu_N \rightarrow \tau_N$.
\end{itemize}
(And, obviously, for each $\tau_i$ there is a related $\sigma_i$.)
Now, amazing enough (but there are mathematical
theorems ensuring the `correct' behavior~\cite{Andrieu}),
the points so obtained {\em sample} the bi-dimensional
distribution $(\mu,\,\tau)$, and then $(\mu,\sigma)$, in the sense that
the expected frequency to visit a given region is proportional
to the probability of that region (just {\em Bernoulli} theorem,
not to be confused with the frequentist `definition' of
probability! -- see e.g. Ref.~\cite{GdA_ProbabilityPropensity}).
Moreover, for the way it has been described, it is clear that the probability
of the move $(\mu_i,\tau_i)\rightarrow (\mu_{i+1},\tau_i)$ depends
only $(\mu_i,\tau_i)$ and \underline{not} on the previous {\em states}.
This is what defines a {\em Markov Chain Monte Carlo}, of which the
Gibbs sample is one of the possible algorithms.

There is still the question of $f_0(\tau)$, less trivial
then $f_0(\mu)$, because $\tau$ has to be
positive.\footnote{Also a mass, as many other physics quantities,
  is positively defined, and in principle one has to
  pay attention,
  either in the sampling steps or when the resulting {\em chain}
  is analyzed,
  that it does not get negative
  But this problem does not occur in practice if the
  the average value $\overline{x}$ is many standard $\sigma_C$
  above zero. Anyway, packages like {\tt JAGS} allow
   also sharp constrains on the priors.
  (This is general problem when we use Gaussians to describe
  positively defined quantities, already realized by Gauss
  and reminded in footnote \ref{fn:Gauss}.)
}
In this case a convenient prior would be a Gamma, with
$\alpha_0$ and $\beta_0$ properly chosen, because
it easy to see that, when multiplied by
Eq.~(\ref{eq:Gamma_tau}), the result is still a Gamma:
\begin{eqnarray}  
  f(\tau\,|\,\underline{x},\mu^*)  &\propto &
  \tau^{\alpha-1}\,e^{-\beta\,\tau} \cdot
  \tau^{\alpha_0-1}\,e^{-\beta_0\,\tau} \\
   &\propto &  \tau^{\alpha+\alpha_0-2}\,e^{-(\beta+\beta_0)\,\tau}\,.
  \label{eq:Gamma_tau_gamma_prior}
\end{eqnarray}
We can easily see that a flat prior for
$\tau$ is recovered in the limit $\alpha_0\rightarrow 1$
and $\beta_0\rightarrow 0$.

A last comment concerning the initial point for the sampling
is in order.
Obviously, the initial steps of the {\em history} (the sequence)
depend on our choice, and therefore they can be somehow
not `representative'. The usual procedure to overcome
this problem consists in discarding
the `first points' of the sequence, better if after
a visual inspection, or using criteria based on past experience
(notoriously, {\em this kind of techniques are between science and art},
even when they are grounded on mathematical theorems, which
however only speak of `asymptotic behavior').
But, as a matter of fact, the convergence of the Gibbs sampler
for low dimensional problems is very fast and modern computers
are so powerful that, in the case of doubt,
we can simply throw away several thousands initial
points `just for security'.

\subsection{Implementation in {\tt JAGS}/{\tt rjags}}\label{sec:IntroJags}
As a first example, here is the model to make
the simple weighted average of the
charged kaon mass values of
Tab.~\ref{tab:masseK_PDG} (we are indeed ``breaking a nut with a mallet''):\\
\vspace{-0.6cm}
{\footnotesize
\begin{verbatim}
model {
   for (i in 1:length(d)) {
      d[i] ~ dnorm(m, 1/s[i]^2);
   }
   m  ~ dnorm(0.0, 1.0E-8);
}
\end{verbatim}
}
\noindent
The loop is just the implementation of the graphical model
on the right side of Fig.~\ref{fig:StandardComb}, with
$x_i$ here called {\tt d[i]} in order to maintain the notation used
in Eqs.~(\ref{eq:media}) and (\ref{eq:sigma}).
{\tt dnorm()} stands for \underline{norm}al distribution 
\underline{d}ensity function, whose parameters are the kaon mass
          {\tt m}
(equal for all measurements, because we believe they
were measuring the same thing) and
{\tt 1/s[i]\^{}2}, that is $\tau_i$, as
discussed in the previous subsection.
The last line of code defines the prior of the mass value: formally
a normal distribution, but in fact a flat one in the domain of
interest, being $\sigma_0=10^4\,$MeV. The model is saved in the
file {\tt weighted\_average.bug} and we move now to the {\tt R} code.

First we assign the experimental values to the vector {\tt d} and
{\tt s} (no declarations are required in {\tt R}) and then we
evaluate
and print the weighed average and the combined standard deviation
calculated from Eqs.~(\ref{eq:media}) and (\ref{eq:sigma}): 
\vspace{0.5cm}
{\footnotesize
\begin{verbatim}
d <- c(493.691, 493.657, 493.670, 493.640, 493.636, 493.696)
s <- c(  0.040,   0.020,   0.029,   0.054,   0.011,   0.007)
d.av  <- sum(d/s^2)/sum(1/s^2) 
s.av  <- 1/sqrt(sum(1/s^2))
cat(sprintf("combined value: %f +- %f\n", d.av, s.av))
\end{verbatim}
}
\noindent
Executing the script\footnote{If these
 lines are saved
  in a file, e.g. {\tt kaon\_mass\_naive.R}, 
  then the script can be run with the command
  {\tt source('kaon\_mass\_naive.R')}.
}
containing these five lines, we get
{\footnotesize
\begin{verbatim}
combined value: 493.676599 +- 0.005478
\end{verbatim}
}
\noindent
that for the moment is just a check.

\newpage
Let us now move to the
{\tt rjags} stuff in the {\tt R} script:
{\footnotesize
\begin{verbatim}
library(rjags)   # load rjags 
data   <- NULL   # declare an empty list
data$d <- d      # first  element of the list
data$s <- s      # second element of the list
model  <- "weighted_average.bug"  

jm <- jags.model(model, data)                    # define the model
update(jm, 100)                                  # burn in 
chain <- coda.samples(jm, c("m"), n.iter=10000)  # sampling 
\end{verbatim}
}
\noindent
So, first the package {\tt rjags} is loaded 
calling the function {\tt library()}, then we fill the data
in the {\em list}\footnote{A `list' is a very
interesting object of {\tt R}, which can contain other objects, also
of different kinds and different lengths; the element of a `list'
can be accessed either by name, as we do here, or by indices.} {\tt data} (arbitrary name) and
we put the model file name into
the string variable {\tt model} (again arbitrary name).  Finally
we interact with {\tt JAGS} in three steps:
\begin{enumerate}
\item the function {\tt jags.model(model, data)} 
  passes model and the data to JAGS; the model is compiled and,
  if everything is ok, a summary message is reported, like in this case
{\footnotesize
\begin{verbatim}
Compiling model graph
   Resolving undeclared variables
   Allocating nodes
Graph information:
   Observed stochastic nodes: 6
   Unobserved stochastic nodes: 1
   Total graph size: 29
\end{verbatim}
}
\item then, {\tt update(jm, 100)} lets the Markov chain do 100 moves
  in the parameter space, but without recording the values,  
  so that the initial points are not
  taken into account when the chain if analyzed;
\item  finally, {\tt coda.samples(jm, c("m"), n.iter=10000)}
  does the real work, following the evolution of the chain
  for {\tt n.iter} steps,
  the model variable {\tt m} is monitored and the resulting history
  is returned and stored in the object {\tt chain} (arbitrary name).
\end{enumerate}  
At this point {\tt JAGS} has done its work and we only need
to analyze its outcome. For this task 
the high level functions of {\tt R} are very helpful.
For example we can make a summary plot just calling {\tt  plot(chain)}, 
\noindent
whose result is shown in Fig.~\ref{fig:naive_chain}:
\begin{figure}[!t]
\begin{center}
\begin{tabular}{c}    
  \epsfig{file=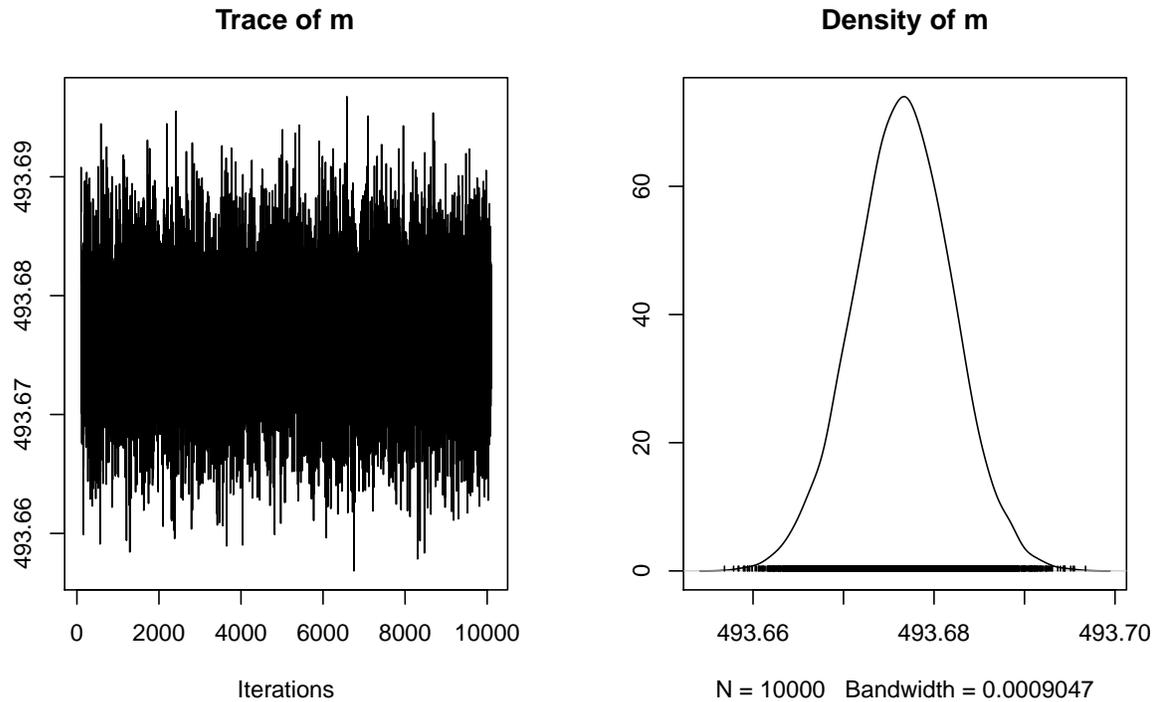,clip=,width=\linewidth}
\end{tabular}  
\end{center}  
\caption{\small \sf Summary plot of the chain returned by JAGS.}
\label{fig:naive_chain}
\end{figure}
the left hand plot shows the history of {\tt m} during the 10000
recorded iterations;
the right hand one is a smoothed representation of the histogram
of the sampled values of {\tt m}.

{\tt R} provides also a summary of the result, using the
function {\tt summary(chain)} $[$or, better, 
{\tt print(summary(chain))} if we want to include it into a script$]$,
with many statistical informations like average,
standard deviation and quantiles
for each sampled variable.
Or we can do it in more detail using the high level
{\tt R} functions.
Here is, for example, how to calculate mean and standard deviation
(also to show a way to extract the history
of a single variable from the object returned by {\tt coda.samples()}):
{\footnotesize
\begin{verbatim}
m.mean <- mean(chain[[1]][,'m'])
m.sd   <- sd(chain[[1]][,'m']) 
cat(sprintf("JAGS result: %f +- %f\n", m.mean, m.sd))
\end{verbatim}
}
\noindent
resulting (with this particular sampling) in
{\footnotesize
\begin{verbatim}
JAGS result: 493.676632 +- 0.005451
\end{verbatim}
}
\noindent
practically identical
to the weighted average obtained using exact formulae.

\section{Sceptical combination with {\tt JAGS} --
 Preliminaries}
It is now time to improve our model in order to implement
what has been discussed in the introduction, where it was said
that our skepticism would act on the variable $r_i = \sigma_i/s_i$.
However, following Ref.~\cite{Dose},
in the previous paper on the subject~\cite{sceptical1999} a different variable
was indeed considered, namely $\omega_i = s_i^2/\sigma_i^2$.
The relations between the  variables which enter the game are
conveniently shown
in the graphical model of Fig.~\ref{fig:model_DL}, where
the following convention has been used:
\begin{figure}[!t]
\begin{center}   
  \epsfig{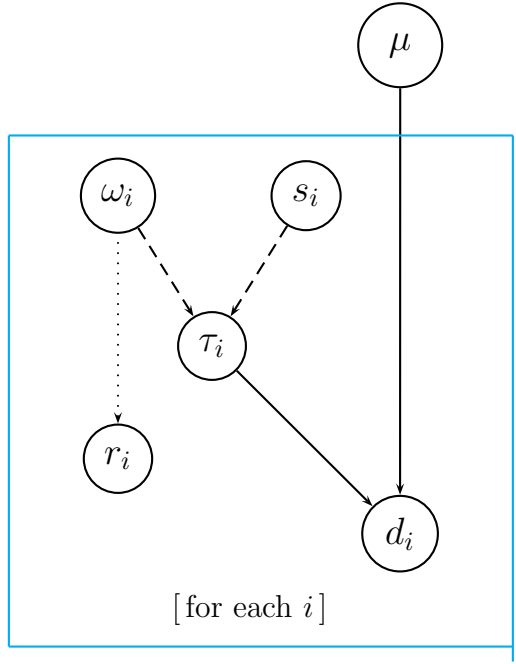} 
\end{center}  
\caption{\small \sf Graphical model behind the sceptical combination.
  The solid lines represent probabilistic links, the dashed ones
  deterministic links. The unusual dotted line stands for an auxiliary
  variable which is not really part of the model, but it is convenient
  to track.
}
\label{fig:model_DL}
\end{figure}
\begin{itemize}
\item arrows with solid lines represent -- this is the usual
  convention --
  {\em probabilistic links} between {\em parent(s)} and {\em child(ren)},
  in our case $d_i$ depending from $\mu$ and $\tau_i$;
\item arrows with dashed lines represent {\em deterministic links}, that is
  $\tau_i = \omega_i/s_i^2$;
\item finally the arrow with {\em dotted line} is {\em unusual}, in the sense
  that it is not used in the literature. It is in fact 
  still a deterministic link, being $r_i = 1/\sqrt{\omega_i}$,
  but it is {\em irrelevant} for the model itself and it could be
  also (and perhaps better, as far as the efficiency of the
  program is concerned) calculated at the end of the sampling. 
\end{itemize}
In the graphical model there are three kind parents having no
`ancestors':
$\mu$, $\omega_i$, and $s_i$. Therefore they need priors,
that is $f(\mu\,|\,I)=f_0(\mu)$, and so on.
But $s_i$ are simply constant
and do not require priors (or, if you like, they are just
Dirac delta's). For $f_0(\mu)$ we choose, as before,
a practically flat prior obtained by a Gaussian distribution with
very large $\sigma_0$. The mathematically convenient prior of $\omega_i$ is
instead a Gamma distribution,
implying that  the distribution of $r_i$ is instead
not an elementary one, as
it can be seen comparing
Eq.~(11) and (12) of Ref.~\cite{sceptical1999}.

\begin{figure}[!t]
\begin{center}
\begin{tabular}{|c|c|}\hline
 Individual results  & Combined results \\ \hline 
\epsfig{file=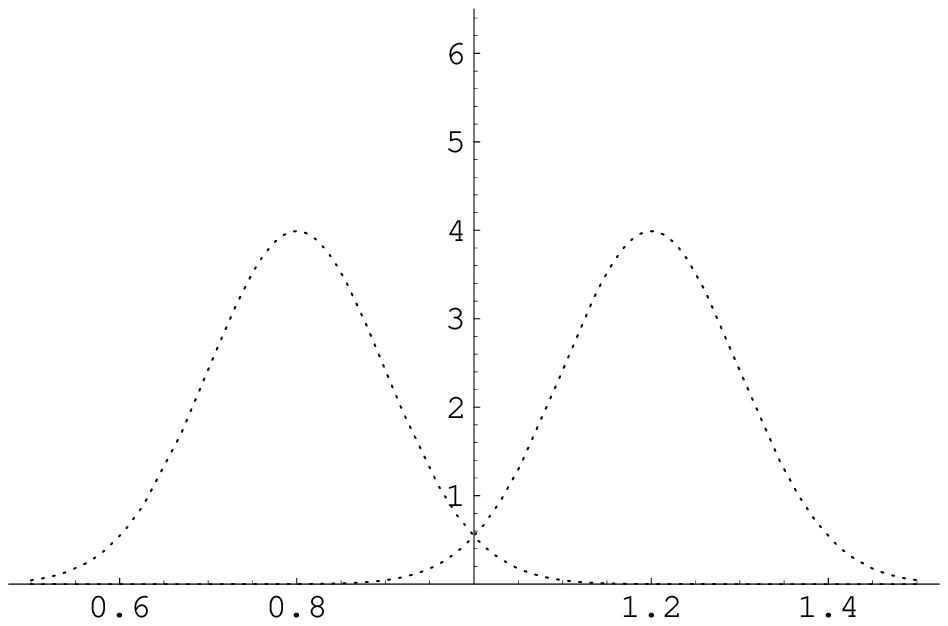,width=0.4\linewidth,clip=}  &
\epsfig{file=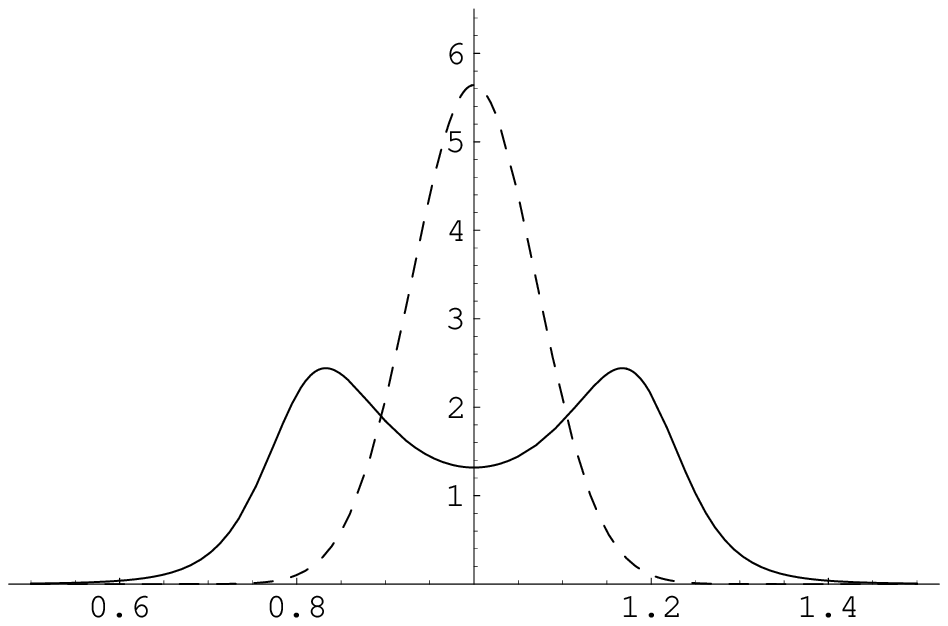,width=0.4\linewidth,clip=}  \\\hline
\epsfig{file=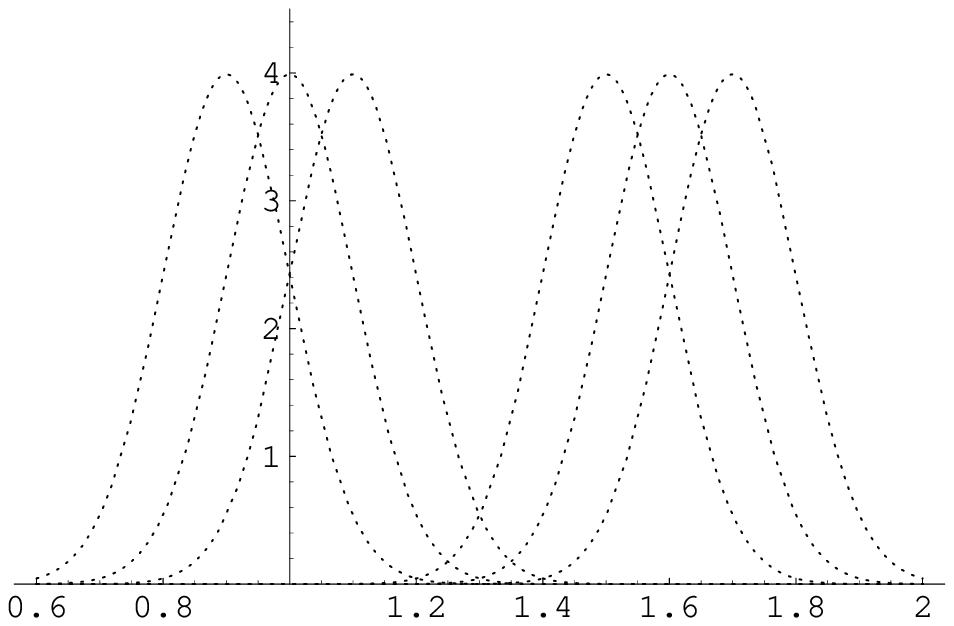,width=0.4\linewidth,clip=}  &
\epsfig{file=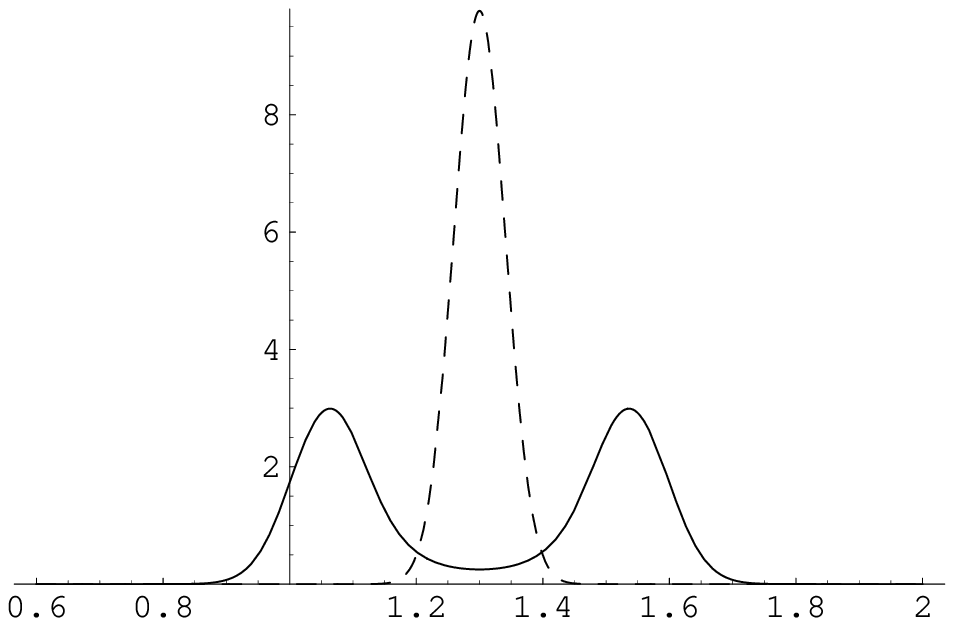,width=0.4\linewidth,clip=} \\ \hline
\epsfig{file=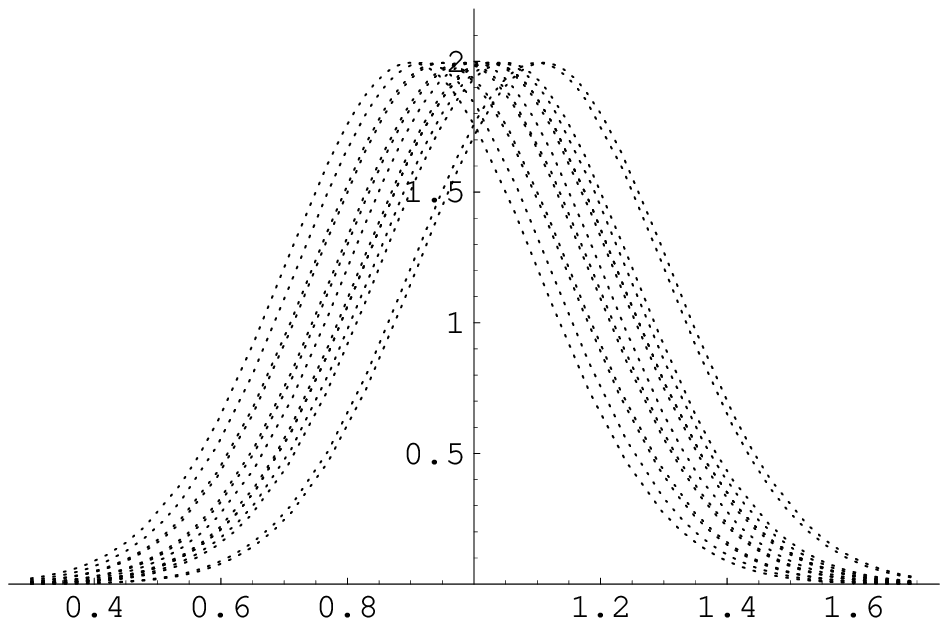,width=0.4\linewidth,clip=}  &
\epsfig{file=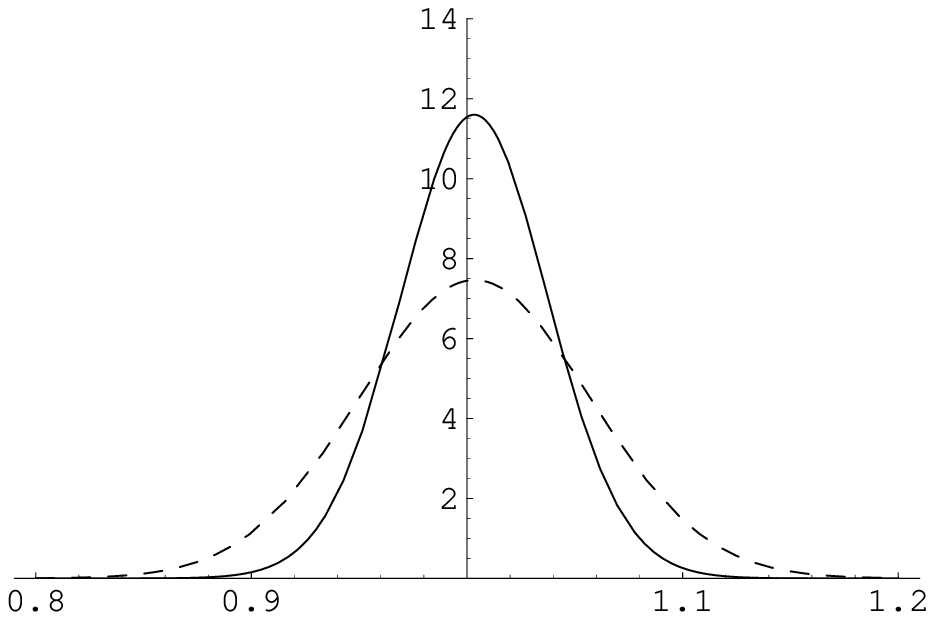,width=0.4\linewidth,clip=} \\ \hline
\end{tabular}
\end{center}
\caption{\small Examples of sceptical combinations
  taken from \cite{sceptical1999}.
The plots on the left-hand side  show the individual results.
The  plots on the right-hand side  show the combined result obtained 
using a sceptical combination (continuous lines), 
compared with the standard combination (dashed lines).} 
\label{fig:from_9910036}
\end{figure}

For the parameters
  of the Gamma pdf
of $\omega_i$ 
we stick to those chosen in Ref.~\cite{sceptical1999},\footnote{For easier
  comparison
  with the results of Ref.~\cite{sceptical1999} 
  for the Gamma parameters we use hereafter
 $\delta$ and $\lambda$ instead of the standard $\alpha$ and $\beta$.} 
i.e. $\delta=1.3$ and $\lambda=0.6$,
in order to get $\mbox{E}[r_i] = \sigma(r_i) = 1$.
Figure ~\ref{fig:from_9910036},
taken from Fig.~4 of
Ref.~\cite{sceptical1999}, shows how the model performs 
in some  situations which represent kind of extreme cases
with respect to the usual `disagreements' within a set of
results.

Particularly interesting are the cases in which 
the individual results cluster in two regions, or when
they overlap `too much'. In the first case, while the simple
weighted average prefers mass values in a region where there is no experimental
support, the sceptical combination exhibits a bimodal distribution,
because we tend to believe with equal probability that the true value
is in either side
 (but it could also be in the middle, although with
low probability). In the second case, instead, in which the results overlap
too much,  the method has the nice feature of
producing a pdf narrower than that obtained by the 
simple weighted average, reflecting our natural suspicion
that the quoted uncertainties might have been
overestimated. In Ref.~\cite{sceptical1999} (Fig.~5 there)
it was also studied how the results varied if the initial
$\sigma(r_i)$ was allowed to move by $\pm 50\%$.
This leads us to be rather 
confident that the choice of  $\delta=1.3$ and $\lambda=0.6$
is not critical.

Here is, finally, the {\tt JAGS} model,
easy to understand if we compare it
with the graphical model of  
Fig.~\ref{fig:model_DL}:
%
{\footnotesize
\begin{verbatim}
var tau[N], r[N], omega[N];
model {
   for (i in 1:N) {
      d[i] ~ dnorm(mu, tau[i]);
      tau[i] <- omega[i]/s[i]^2;
      omega[i] ~ dgamma(delta, lambda);
      r[i] <- 1.0/sqrt(omega[i]);
   }
   mu  ~ dnorm(0.0, 1.0E-8);
}
\end{verbatim}
}
\noindent

\vspace{0.2cm}\mbox{}

Before running it, let us make a very trivial model
in which JAGS is used as a simple random generator, without
any inferential purpose, just to get confidence with the
prior distribution of $r_i$. Moreover, consisting
the core of the model of just two lines of code,
we write it directly from the {\tt R} script into a temporary file.
Here is the complete script, in which it also shown  an
alternative way (indeed the simplest one in {\tt R})
to prepare the `list' {\tt data} to be passed to {\tt JAGS}
via {\tt jags.model()} -- note  the missing {\tt update()}, because
we deal here with direct sampling and there are no burn-in issues:
\newpage
{\footnotesize
\begin{verbatim}
library(rjags)
data = list(delta=1.3, lambda=0.6)
model = "tmp_model.bug"
write("
model{
  omega ~ dgamma(delta, lambda);
  r <- 1.0/sqrt(omega);
}
", model)

jm <- jags.model(model, data)
chain <- coda.samples(jm, c("omega","r"), n.iter=10000)
plot(chain)
print(summary(chain))
\end{verbatim}
}
\noindent
This is the result of the last command:
{\footnotesize
\begin{verbatim}
1. Empirical mean and standard deviation for each variable,
   plus standard error of the mean:

        Mean     SD Naive SE Time-series SE
omega 2.1979 1.9127 0.019127       0.019065
r     0.9948 0.9096 0.009096       0.009096

2. Quantiles for each variable:

        2.5%    25%    50%   75% 97.5%
omega 0.1162 0.8045 1.6582 3.035 7.329
r     0.3694 0.5741 0.7766 1.115 2.934
\end{verbatim}
}
\noindent
We see that the (indirectly) sampled {\tt r} has
(with good approximation)
the expected  unitary mean and standard deviation.
As a further check, let us use directly the Gamma random
generator {\tt rgamma()} of {\tt R},
{\footnotesize
\begin{verbatim}
r <- 1/sqrt(rgamma(10000, 1.3, 0.6))
cat(sprintf("mean(r) = %f, sd(r) = %f\n", mean(r), sd(r)))
\end{verbatim}
}
\noindent
whose (aleatory) result is left as exercise to the reader.

\section{Sceptical combination with {\tt JAGS} --
  Results}\label{sec:JAGSresults}
\begin{figure}
  \begin{center}
    \begin{tabular}{c} 
      \epsfig{file=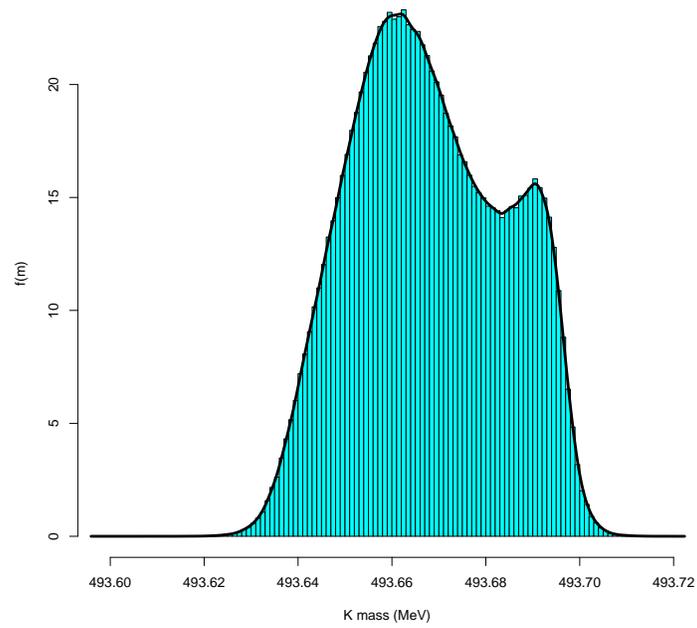,clip=,width=0.6\linewidth} \\
 \mbox{} \\
      \epsfig{file=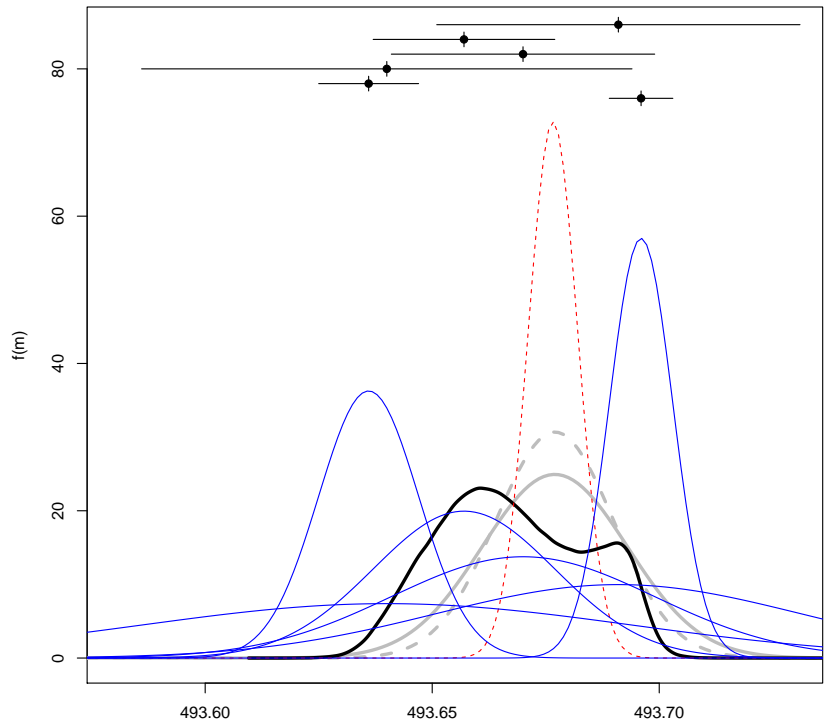,clip=,width=0.75\linewidth}
  \end{tabular}   
  \end{center}
  \caption{\small \sf Above: histogram of charged kaon mass by MCMC sampling.
    Below: profile of the above histogram compared with the
    individual results, the naive weighted average (dashed) and
    the Gaussian based on the average and the PDG prescription to scale the
    `error' (see text).
}
\label{fig:sceptical_m}
\end{figure}

The core of the {\tt R} script is very similar to the one
of section
\ref{sec:IntroJags}, besides the model used, the number
of iterations and the variables to be monitored.
Note that we pass the Gamma parameter {\tt delta}
and {\tt lambda} to the model via the list {\tt data}
(a different choice could have been to define them directly inside the model).
Here is the entire script, including statistics and plot summaries
(plots not shown here).
{\footnotesize
\begin{verbatim}
library(rjags)
data <- NULL             # 'data' to be passed to the model
data$d  <- c(493.691, 493.657, 493.670, 493.640, 493.636, 493.696)
data$s  <- c(  0.040,   0.020,   0.029,   0.054,   0.007,   0.007)
data$N  <- length(data$d)
data$delta  = 1.3   #  parameters of Gamma(omega):  delta  <--> 'alpha'  
data$lambda = 0.6   #                               lambda <--> 'beta'

jm <- jags.model(model, data)                            # define the model
update(jm, 1000)                                         # burn in 
chain <- coda.samples(jm, c("mu", "r"), n.iter=100000)   # sampling 

print(summary(chain))
plot(chain)
\end{verbatim}
}
\noindent
This is what we get from {\tt summary(chain)}:
{\footnotesize
\begin{verbatim}
1. Empirical mean and standard deviation for each variable,
   plus standard error of the mean:

         Mean      SD  Naive SE Time-series SE
mu   493.6678 0.01608 1.608e-05      0.0000515
r[1]   0.8790 0.53138 5.314e-04      0.0006477
r[2]   0.9671 0.63559 6.356e-04      0.0010050
r[3]   0.8330 0.49631 4.963e-04      0.0005264
r[4]   0.8450 0.50271 5.027e-04      0.0005771
r[5]   2.1528 1.62050 1.621e-03      0.0034153
r[6]   2.9148 2.35537 2.355e-03      0.0052502

2. Quantiles for each variable:

         2.5%      25%      50%      75%   97.5%
mu   493.6394 493.6555 493.6666 493.6807 493.696
r[1]   0.3763   0.5713   0.7457   1.0179   2.177
r[2]   0.3803   0.5965   0.8054   1.1338   2.515
r[3]   0.3656   0.5467   0.7090   0.9619   2.045
r[4]   0.3703   0.5550   0.7194   0.9753   2.077
r[5]   0.5391   1.1500   1.7672   2.6621   6.101
r[6]   0.5073   1.3988   2.4184   3.7211   8.600
\end{verbatim}
}
\noindent
The result is then a mass of
$493.668\pm 0.016\,$MeV, where the `error' is provided
by the {\em standard uncertainty}~\cite{ISO,ISOD}. However, this 
does not imply that the value of the mass is normally distributed
around the average.\footnote{An example in which the sceptical
  combination produces a result narrower that the
  weighted average is shown in the Appendix.}
This can be checked on the
quantiles of `{\tt mu}' (see above output), and better on the histogram
of the sampled values, shown in the upper plot of 
Fig.~\ref{fig:sceptical_m}, 
whose smoothed profile is reported
in the bottom plot of the
same figure,\footnote{The histogram with the over-imposed profile was produced
  by
  {\footnotesize
    \mbox{}\vspace{-0.4cm}
\begin{verbatim}
chain.df <- as.data.frame( as.mcmc(chain) )
hist(chain.df$mu,nc=100,prob=TRUE,xlab='K mass (MeV)',ylab='f(m)', col='cyan',main='')
lines(density(chain.df$mu, adjust=1.0), lwd=3)
\end{verbatim}
}
}
together with the individual measurements,
the standard weighted average (red dashed) and the combination
got following the PDG prescriptions (gray) with
the $\times 2.4$ (narrower curve) and $\times 2.8$
scaling (wider curve) \cite{PDG2019}.

Here is a summary of the results:\\ \mbox{} \\
\begin{tabular}{ccc}
  method  && $m_{K^{\pm}}$ (MeV)  \\
 \hline 
  simple weighted average && $493.677 \pm 0.006$  \\
  PDG (`OUR AVERAGE', $\times 2.4$ scaling) && $493.677 \pm 0.013 $   \\
  PDG (`OUR FIT', $\times 2.8$ scaling) && $ 493.677\pm 0.016$  \\
  sceptical && $493.668\pm 0.016\,$ \\
  & &
\end{tabular}
 
\mbox{}  \\ 
As far as the rescaling factor $r_i$ are concerned,
we see from the output of
the summary that only those relative to the items 5 and the 6
are preferred
to be higher the the initial ones,
with mean values 2.2 and 2.9, respectively.
Therefore with this data set the most 'suspicious' one is nr. 6,
as we had judged by eye in the introduction. But since the $r_i$
were inferred simultaneously, and together with the mass, it is
interested to give a look to the correlation matrix. Here is directly the
{\tt R} output:
{\footnotesize
\begin{verbatim}
        mu  r[1]  r[2]  r[3]  r[4]  r[5]  r[6]
mu    1.00 -0.21  0.31 -0.03  0.16  0.55 -0.59
r[1] -0.21  1.00 -0.05  0.02 -0.03 -0.11  0.13
r[2]  0.31 -0.05  1.00  0.03  0.06  0.18 -0.17
r[3] -0.03  0.02  0.03  1.00  0.00 -0.01  0.02
r[4]  0.16 -0.03  0.06  0.00  1.00  0.09 -0.09
r[5]  0.55 -0.11  0.18 -0.01  0.09  1.00 -0.32
r[6] -0.59  0.13 -0.17  0.02 -0.09 -0.32  1.00
\end{verbatim}
}
\noindent
We see, for example, that the highest correlation of the mass (`{\tt mu}')
is with {\tt r[5]} and {\tt r[6]}, related
to the two most precise measurements: the first is positive, meaning that
a larger $\sigma_5$ would allow {\tt mu} to rise towards $d_6$;
the second is negative, meaning 
that a larger $\sigma_6$ would allow {\tt mu} to descend towards $d_5$.
For this reason, among the several $r_i$, {\tt r[5]} and {\tt r[6]}
get the highest (in absolute value)
correlation coefficient, having a negative sign. But it is only $-32\%$,
indicating that  $\sigma_5$ and $\sigma_6$ could possibly be both larger
than the stated standard uncertainty.
 
Going back to the mass value, we see that our result does not differ
much from the PDG one, if we are
only interested in average and standard uncertainty
(just $-9\,$keV lower, with similar uncertainty).
What differs mostly is the shape
of the probability distribution, which is
has nothing to do with a Gaussian. Instead, in the the
weighted average, with the resulting `error'
as it comes straight from Eq.~(\ref{eq:sigma})
or scaled with $\sqrt{\chi^2/\nu}$,
the interpretation is {\em tacitly} Gaussian,
or it is assumed as such in further analyses~\cite{GdA_combination_no_sufficiency}. 
For example, if one is
interested, for some deep physical reasons,
in the chance that the mass is
larger than 493.70\,MeV,
it is clear from the bottom plot of Fig.~\ref{fig:sceptical_m}
that the results would be quite different.

\begin{figure}[!t]
  \begin{center}
      \epsfig{file=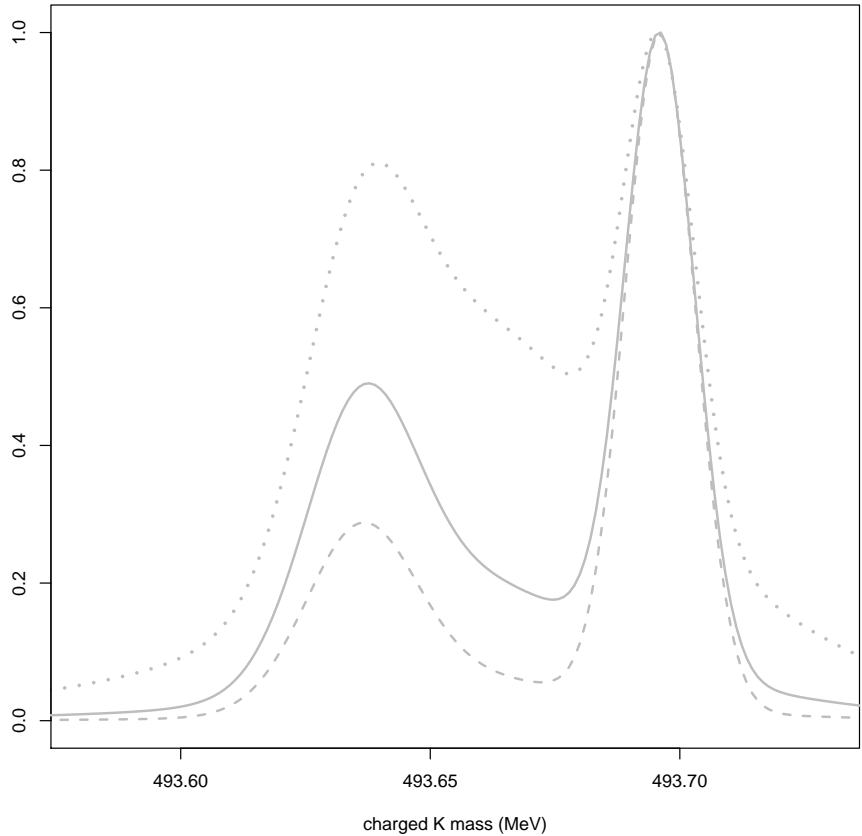,clip=,width=0.75\linewidth} \\
\end{center}  
  \caption{\small \sf Attempts to re-obtain the bi-modal curve shown
    in the PDG {\em ideogram} of Fig.~\ref{fig:PDF2019}. They are not normalized
    but just equalized to the value of their maximum. Solid line:
    $\propto\sum_i f_{\cal N}(m\,|\,d_i,s_i)/s_i$;
    dotted line:   $\propto\sum_i f_{\cal N}(m\,|\,d_i,s_i)$; 
    dashed line:  $\propto\sum_i f_{\cal N}(m\,|\,d_i,s_i)/s_i^2$.
}
\label{fig_PDG_ideogram}
\end{figure}

One might argue if, for such a 
purpose, we could use the bi-modal curve of the PDG
{\em ideogram} (see Fig.~\ref{fig:PDF2019}),
in alternative to the pdf resulting from
the sceptical analysis performed here.
But what is the meaning of the  bi-modal curve of Fig.~\ref{fig:PDF2019}?
If one compares it with the individual Gaussians reported
in Fig.~\ref{fig:NaiveCombination} we see that  it follows somehow
the profile of highest points of the curves. Therefore 
the first guess is that it is just an unnormalized sum, that is
$\sum_if_{\cal N}(m\,|\,d_i,\sigma_i)$. But
checking it, it does not seem to be the case. The second
guess was a kind of weighted average, with weights equal to
$1/s_i^2$, i.e. $\sum_if_{\cal N}(m\,|\,d_i,\sigma_i)/s_i^2$,
but it did not work either. The third attempt was to set the
weights to $1/s_i$, i.e. $\sum_if_{\cal N}(m\,|\,d_i,\sigma_i)/s_i$,
and this seems to be the case.
The three attempts are reported in Fig.~\ref{fig_PDG_ideogram},
but just as a curiosity, as {\em they have no probabilistic meaning}.

Let us end this section showing how to re-obtain
the standard combination with
the same  general model used in the sceptical combination.
We just need to choose values  
suitable $\delta$ and $\lambda$ to get
$\mbox{E}[r_i]=1$ and  $\sigma(r_i)\rightarrow 0$.
Inverting Eqs. (13) and (14) of \cite{sceptical1999} seems complicate,
but in the limit of zero variance, this is the same as requiring
$\mbox{E}[\omega_i]=1$ and $\sigma(\omega_i)\rightarrow 0$,
a condition easier to apply in practice. Being in fact
$\mbox{E}[\omega_i]=\delta/\lambda$ and
$\mbox{var}[\omega_i]=\delta/\lambda^2 = (\delta/\lambda)/\lambda$,
the requirement simply translates into $\delta=\lambda$
with both parameters `very large'. In practice is is enough to set e.g. 
$\delta=\lambda=10000$ to recover the result of the weighed average
of $493.6766\pm 0.0055\,$MeV.

\section{Further scepticism}
If the purpose of this paper would have been just to search around
for a case of `apparent' discordant results,
as a real life example to which
apply to the  model of Ref.\cite{sceptical1999} implemented in {\tt JAGS},
then 
the game would be at the end. But since I am presently interested
in the charged kaon mass, I tried to understand the results a bit more.
I expected in fact to find in the publications
extensive discussions on the details of the analysis,
with explanations of what the results really meant and detailed
accounts for the sources of uncertainties, as it has presently 
become a good practice by most experimental teams. But this
was not the case.
Already trying to understand the (apparently, as we shall see)
most precise value,  I was quite surprised when I
 realized that Ref.~\cite{Kmass91} gives no detailed information on
  how they got their numbers and on what their `error' really means.
  Furthermore the PDG uses an `error' of 0.007\,MeV,
  instead of the 0.0059\,MeV reported by  \cite{Kmass91},
  on the basis of a PhD thesis~\cite{Kmass92}
  which it is impossible to find (not even in Russian!).
  Fortunately this is more
  a methodological paper then a real attempt to get a deep understanding
  of the charged kaon mass, for which a throughout
  analysis of all relevant published
  matter on the subject would be required.\footnote{Presently the
    value of the charged kaon mass, with relative
    uncertainty of around 26\,ppm, is not critical for
    fundamental issues. 
     For example its contribution
  to $|V_{us}|$ of the Standard Model is of the order of 66\,ppm,
  to be combined
  in quadrature with the relative uncertainties of the other quantities
  from which $|V_{us}|$ depends (the branching ratios of interest
  depend on $M_{K^\pm}^5\cdot|V_{us}|^2$ and hence the relative uncertainty
  on $M_{K^\pm}$ is propagated with a factor $5/2$ into the
  relative uncertainty on $|V_{us}|$).
  }

\begin{figure}
  \begin{center}
  \epsfig{file=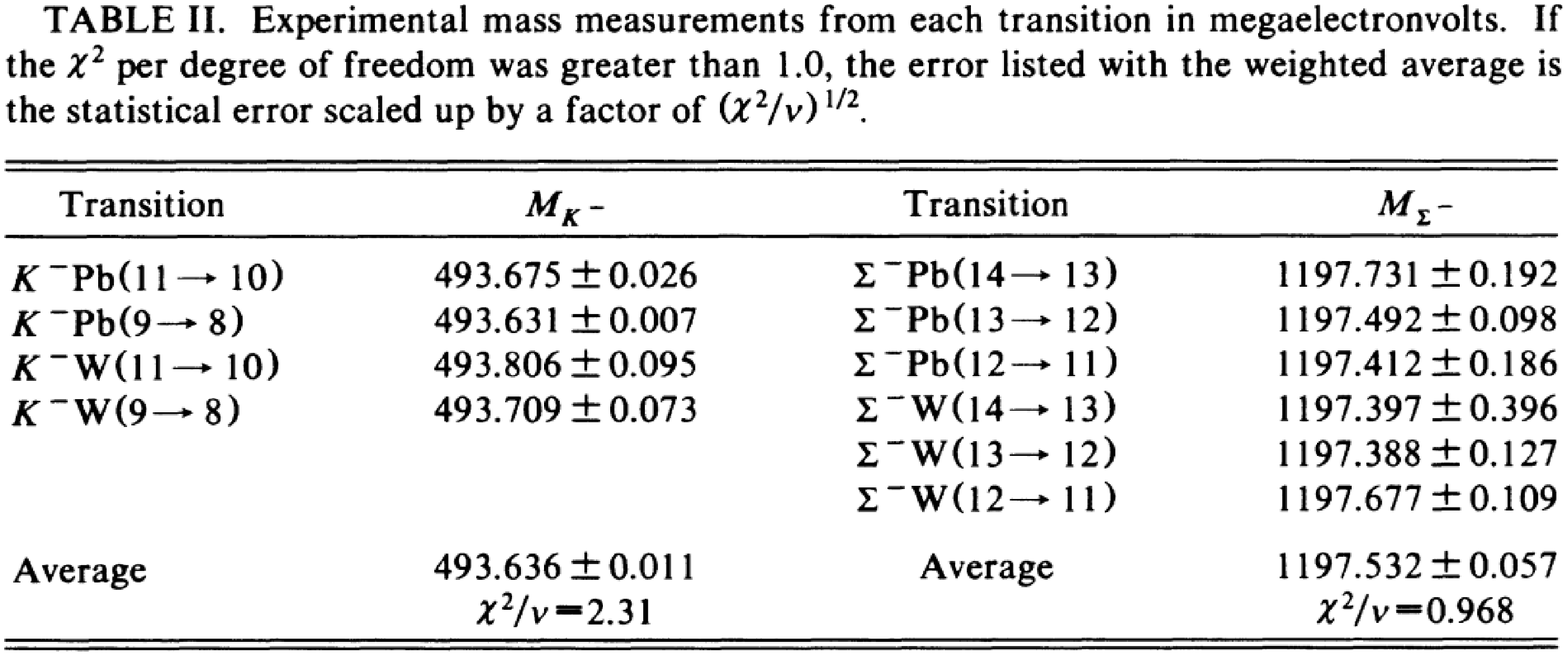,clip=,width=\linewidth}
  \end{center}
  \caption{\small \sf Details of Ref.~\cite{Kmass88}.}
  \label{fig:GallDetails}
  \end{figure}

  Nevertheless, there is a point I would like to touch, related
  to the second most precise result of the list~\cite{Kmass88}
  whose conservative uncertainty is uncritically accepted by the PDG. 
  The paper provides in fact four mass values, reported for
  the reader's convenience
  in Fig.~\ref{fig:GallDetails}. 
  The weighted average is $493.6355 \pm 0.0067$, rescaled (and rounded)
  to $493.636 \pm 0.011$ based on a `high $\chi^2$', which is in reality not so bad, being 
  7.0 with 3 degrees of freedom,\footnote{Providing just `$\chi^2/\nu$'
    is, as now well understood, rather misleading, because
    the $\chi^2$ does not scale with $\nu$. Therefore, though a ratio of 2.31
    would be a clear alarm bell for $\nu=100$, it is quite `in the norm'
    for $\nu=3$.
  } and thus yielding a {\em p-value} of 0.072, above even the (in-)famous
  threshold of 0.05\,\cite{WavesSigmas}. 
  For this reason I could not resist to make a couple of exercises:
  first to see what a sceptical analysis would suggest if we stick
  to the simple weighted average, without the $\sqrt{2.31}\ (=1.52)$ scaling;
  second to see what we get if we make an overall sceptical analysis in  
  which individual results are used.

\subsection{Sceptical analysis using the unscaled result of Ref.~\cite{Kmass88} } 
\begin{figure}[!t]
  \begin{center}
      \epsfig{file=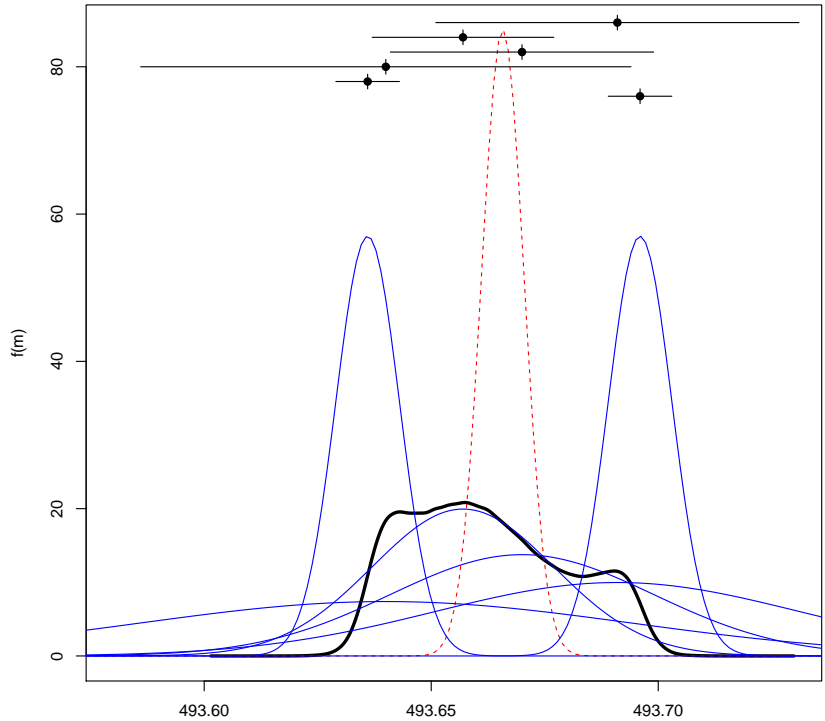,clip=,width=0.75\linewidth}
\end{center}  
  \caption{\small \sf Same as Fig.~\ref{fig:sceptical_m} removing
    the $\times \sqrt{2.31}\ (=1.52)$ scaling applied
    by the authors \cite{Kmass91} to the `error' of the result nr.~5
    of Tab.~\ref{tab:masseK_PDG}.
}
\label{fig:hypersceptical_1}
\end{figure}

Once the  $\times 1.52$ scaling factor on the lowest value is removed,
the weighted average is shifted down and falls right in the middle
of the two most measurements, as shown by the dashed line of
Fig.~\ref{fig:hypersceptical_1},
yielding $m_{K^\pm} = 493.666 \pm 0.005\,$MeV.
But a sceptical analysis yields a broader distribution,
overlapping the two precise measurements on the sides,
thus taking into serious account also the results in between.
Having to report the result as average and standard deviation of the
distribution we get then $493.662 \pm 0.017\,$MeV
(but remember that the complete result is provided by the posterior
pdf evaluated by MCMC sampling).

Looking into the details of the inference, we see that, as imaginable,
high values for $r_5$ and $r_6$ are preferred ($2.8\pm 2.4$ and
$3.4\pm 2.7$, respectively), while the others remain more or less
around the prior values of $\approx 1.0$.

\subsection{Sceptical analysis using the individual values of  Ref.~\cite{Kmass88} }  
Let us know go into the details of the results which contribute
to 5-th entry of Tab.~\ref{tab:masseK_PDG}, reported
in Fig.~\ref{fig:GallDetails}~\cite{Kmass88}
and in the entries 5-8 of Tab.~\ref{tab:masseK_PDG_Gall}.
\begin{table}[t]
\begin{center}
{\footnotesize
  \begin{tabular}{|c|l|c|c|c|c|}
    \hline
 &   \multicolumn{1}{|c|}{Authors} & pub. year &  $[d_i]$ & $[s_i]$\\
 $i$            &        &   &    (MeV)      &    (MeV)    \\
    \hline
 $1$ &  G. Backenstoss et al. \cite{Kmass73}   & 1973 &  493.691 & 0.040 \\
 $2$ &     S.C. Cheng et al. \cite{Kmass75}        & 1975 & 493.657  & 0.020 \\
 $3$ &     L.M. Barkov et al.\cite{Kmass79}       & 1979 &  493.670  &  0.029 \\
    $4$ &     G.K. Lum et al. \cite{Kmass81}       & 1981   &  493.640  &   0.054 \\
\hline    
$5$ &     K.P. Gall et al. \cite{Kmass88}      &  1988  & 493.675 &  0.026   \\
$6$ &                                          &        & 493.631 &  0.007   \\
$7$ &                                          &        & 493.806 &  0.095   \\
$8$ &                                          &        & 493.709 &  0.073   \\
\hline   
    $9$ &     A.S. Denisov et al. \& Yu.M. Ivanov~\cite{Kmass91,Kmass92}   &
    1991  & 493.696  &   0.007  \\
  \hline
  \end{tabular}
  }
\caption{\small \sf Details of the individual result used in the overall analysis. 
}
\label{tab:masseK_PDG_Gall}
\end{center}
\end{table}
\begin{figure}[!b]
  \begin{center}
      \epsfig{file=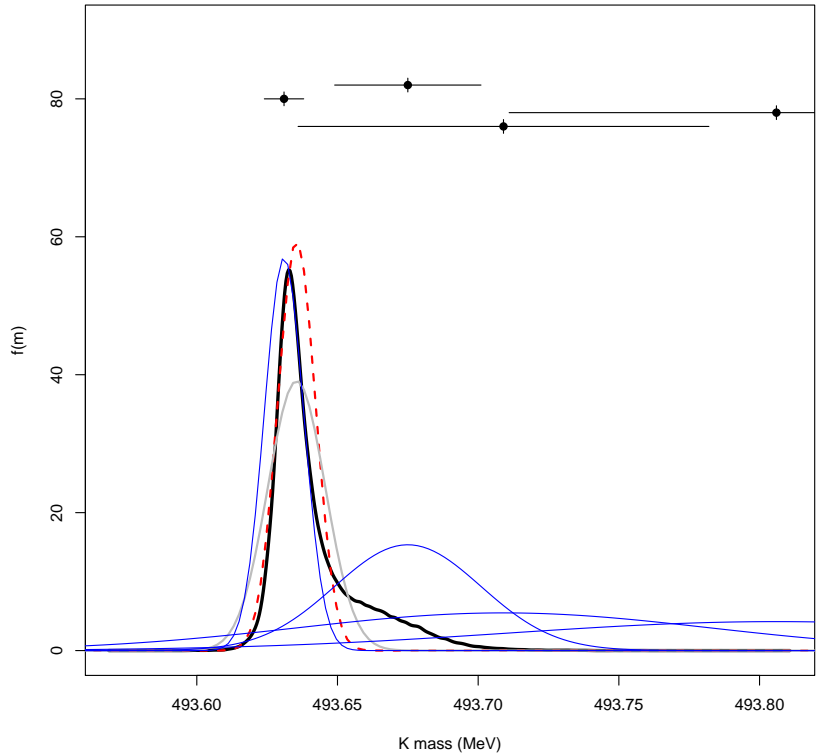,clip=,width=0.71\linewidth}
\end{center}  
  \caption{\small \sf
    Standard and sceptical combination of the four data points
    of Ref.~\cite{Kmass88}. The dashed line is the simple weighted average;
    the gray line having the same center and larger width is the
    same with the standard deviation scaled by $\sqrt{\chi^2/\nu}$
    as done in Ref.~\cite{Kmass88}. The thick, asymmetric curve represent
    the result of the sceptical combination.
}
\label{fig:sceptical_gall}
\end{figure}
There is one high precision value
favoring a small mass value ($493.631\pm 0.007\,$MeV),
and three values of minor precision preferring higher
mass values.
The simple weighted average
of $493.6355\pm 0.0067\,$MeV is then practically equal
to the highest precision value. But then a 
$\times 1.52$ scaling is applied by the authors.
The combined uncertainty grows up,
which is something desirable, but 
it does it symmetrically around the mean,
not taking into account the fact that the other
results would pull the mass value up.

It is then interesting to make a sceptical combination of these
four points. The result is shown if Fig.~\ref{fig:sceptical_gall}.
The sceptical analysis takes into account also the results favoring
higher mass values, although the peak of the distribution (the `mode')
remains very close to the most precise result,
and there is a substantial overlap with it.
The distribution is now skewed on the right side, assigning
higher probability that the mass value is, for example, above
$493.65\,$MeV
with respect to what we could think judging from
mean and standard deviation alone. The resulting mass is
$493.642\pm 0.016\,$MeV shifted up by about $6\,$keV, with a standard
uncertainty about 50\% larger than that provided 
by the $\sqrt{\chi^2/\nu}$ scaling prescription.
But what is more interesting is that the latter
(gray line in the figure,
just below the red dashed one) does not give a correct account of the
possible values of $m_{K^\pm}$, because: {\em i}) it is extended to the
low mass values sizable more than the measured points would allow it;
{\em ii}) it gives practically no chance to mass values above e.g.
$493.657\,$MeV.

Finally there is the question of combining this result with the
other five ones of other experiments. What should we use as input
for the global analysis? Honestly, at this point
we cannot pretend to have not seen the outcome shown in 
Fig.~\ref{fig:sceptical_gall} and to use just the resulting average and
standard uncertainty. We also cannot feed into the model
the complicated posterior we have got. Therefore the only
solution is to make a new combined analysis, but using
all individual results of Ref.~\cite{Kmass88}.
For sake of clarity all points are repeated in
Tab.~\ref{tab:masseK_PDG_Gall}.
\begin{figure}
  \begin{center}
 \epsfig{file=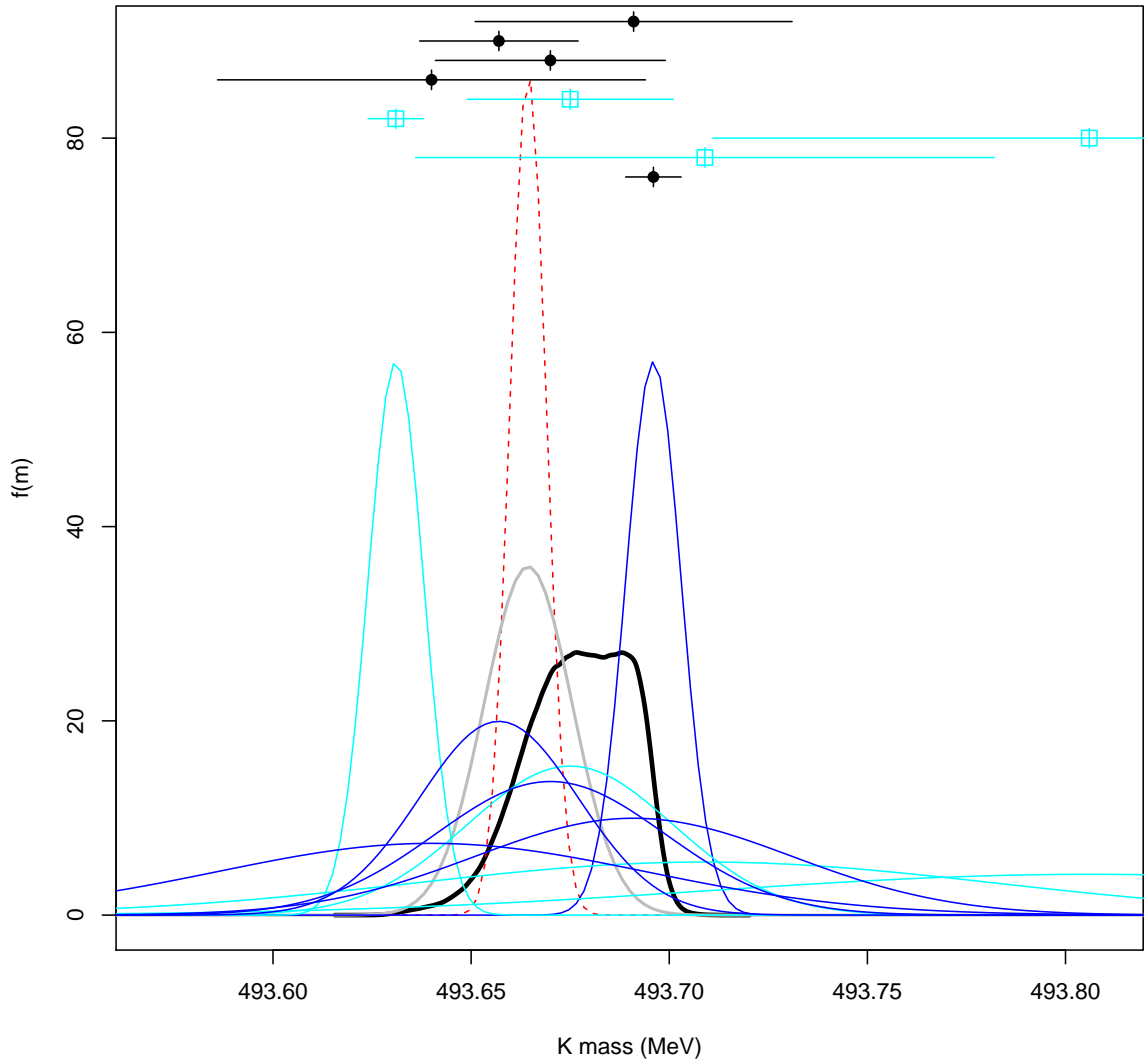,clip=,width=\linewidth}   
  \end{center}
  \caption{\small \sf Combined analysis obtained considering the
    individual points of \cite{Kmass88}. The not trivial final
    pdf of the sceptical analysis (thick continuous black line)
    can be {\em summarized} as $493.677\pm 0.013\,$MeV. The weighed
    average (dashed red) leads instead $493.6644\pm 0.0046$,
    which becomes  $493.664\pm 0.011$ (solid gray Gaussian
    just below the dashed red one) when the standard deviation
    is scaled by the factor $\sqrt{\chi^2/\nu} = \sqrt{47.7/8} = 2.42$.
  }
%
\label{fig:hyper_sceptical_a}
\end{figure}
The result of the analysis, plotted
 in Fig~\ref{fig:hyper_sceptical_a}, is quite surprising
on a first sight: while the standard weighted
average is practically the same
of Fig.~\ref{fig:hypersceptical_1}
(small differences might be attributed to
rounding\footnote{Indeed, if we just calculate  weighted
  averages and related standard deviations, with no
  arbitrary scaling, the result does not change if we use
  the individual results or we group them in steps.
  This is related to the important concept of `statistical
  sufficiency', that will be treated in detail, for the Gaussian case,
  in the forthcoming Ref.~\cite{GdA_combination_no_sufficiency}.}),
the sceptical
combination moves up, disfavoring the low mass solution and
yielding $493.677\pm 0.013\,$MeV.

{\em ``The same as the PDG result''}, 
one would promptly shout at this point,
{\em ``and after so much work!''}. Well, yes and no\ldots\  
Indeed, the PDG numbers
were obtained considering an {\em arbitrarily enlarged}
uncertainty for the combined result of Ref.~\cite{Kmass88}.
Applying, instead, the scaling prescription  to
the nine individual points of Tab.~\ref{tab:masseK_PDG_Gall}
a value of  $493.664\pm 0.011\,$ MeV would have been obtained,
$13\,$keV below the result of the sceptical analysis
(see Fig.\ref{fig:hyper_sceptical_a}).
Certainly this $\approx - 1\,\sigma$ bias will not harm
our understating of fundamental physics, but it is better
to avoid this kind of biases because they could 
perhaps be important in other measurements. 
\subsubsection{Correlations among the values jointly inferred in
the global sceptical analysis}
Let us conclude showing further details concerning the
MCMC sampling.
Here is the output of the {\tt R} command
{\tt summary(chain)} concerning the ten
model parameters:\footnote{Someone would be surprised
  about the possibility of inferring a number of parameters
  superior to the number of the data points. This is not really
  a conceptual problem, as long as we understand that
  they are correlated, often in a complicate way and of which
  the correlation matrix is just a first order representation
  (and we have to be careful when using it in further
  analyses~\cite{CovarianceMatrix}).
}  
{\footnotesize
\begin{verbatim}
1. Empirical mean and standard deviation for each variable,
   plus standard error of the mean:

         Mean      SD  Naive SE Time-series SE
mu   493.6769 0.01285 1.285e-05      3.564e-05
r[1]   0.8101 0.48421 4.842e-04      5.354e-04
r[2]   1.0698 0.69051 6.905e-04      9.878e-04
r[3]   0.8171 0.48464 4.846e-04      5.161e-04
r[4]   0.8903 0.53044 5.304e-04      5.713e-04
r[5]   0.8175 0.49282 4.928e-04      5.288e-04
r[6]   4.5338 3.03359 3.034e-03      4.704e-03
r[7]   1.1908 0.70179 7.018e-04      7.371e-04
r[8]   0.8121 0.48643 4.864e-04      5.035e-04
r[9]   2.0938 1.76421 1.764e-03      3.675e-03

2. Quantiles for each variable:

         2.5%      25%      50%      75%   97.5%
mu   493.6497 493.6680 493.6778 493.6872 493.697
r[1]   0.3588   0.5322   0.6889   0.9332   1.986
r[2]   0.4098   0.6661   0.9014   1.2591   2.736
r[3]   0.3617   0.5381   0.6958   0.9420   1.999
r[4]   0.3928   0.5865   0.7584   1.0274   2.175
r[5]   0.3608   0.5370   0.6954   0.9426   2.007
r[6]   1.3815   2.7852   3.8487   5.4038  11.809
r[7]   0.5333   0.7871   1.0145   1.3696   2.905
r[8]   0.3644   0.5374   0.6922   0.9338   1.979
r[9]   0.4660   0.9894   1.6476   2.6319   6.396
\end{verbatim}
}
\noindent
As we can see, at this point the most suspicious measurement is the 6-th of
the complete list, as we can better judge from the quantiles indicating
that, for example, there is only about 2.5\% probability
that $r_6$ is below $1.38$.
\newpage
Let us also give a look at the correlation matrix:\footnote{Technical remark:
  the correlation matrix has been obtained by the {\tt R} function {\tt cor()},
  applied
  to the chain after a suitable transformation.
  For example, one can transform it into a {\em data frame} and then
  apply {\tt cor()} to it:
{\footnotesize
\begin{verbatim}  
> chain.df <- as.data.frame( as.mcmc(chain) )
> round(cor(chain.df),2)             
\end{verbatim}
}
\noindent  
that includes the rounding at two decimal digits ('$>$' is the {\tt R} prompt).

Or, more simply, we can convert the chain into a matrix,
each column containing the occurrences of each variable during the sample,
and calculate then the correlations between them. This is
how to do it in short, with nested calls to functions
(remember also {\tt print()}, if the command hat to be included into a script):
{\footnotesize
\begin{verbatim}  
> round(cor(as.matrix(chain)),2)   
\end{verbatim}
}
And here are some useful commands to understand what is going on:
{\footnotesize
\begin{verbatim}  
> chain.M <- as.matrix(chain)
> str(chain.M)
> dimnames(chain.M)
> mean(chain.M[,"mu"])
> mean(chain.M[,1])
> mean(chain.M[,"r[9]"])
> mean(chain.M[,10])
> cor(chain.M[,"mu"], chain.M[,"r[9]"])
> cor(chain.M[,1], chain.M[,10])
\end{verbatim}
}
}

{\footnotesize
\begin{verbatim}
        mu  r[1]  r[2]  r[3]  r[4]  r[5]  r[6]  r[7]  r[8]  r[9]
mu    1.00 -0.15  0.33  0.07  0.15 -0.01  0.40 -0.11 -0.10 -0.61
r[1] -0.15  1.00 -0.04  0.00 -0.02  0.01 -0.06  0.02  0.02  0.10
r[2]  0.33 -0.04  1.00  0.04  0.05  0.02  0.14 -0.03 -0.03 -0.20
r[3]  0.07  0.00  0.04  1.00  0.01  0.02  0.03 -0.01  0.00 -0.03
r[4]  0.15 -0.02  0.05  0.01  1.00  0.00  0.06 -0.02 -0.01 -0.09
r[5] -0.01  0.01  0.02  0.02  0.00  1.00  0.00  0.00  0.00  0.02
r[6]  0.40 -0.06  0.14  0.03  0.06  0.00  1.00 -0.04 -0.04 -0.25
r[7] -0.11  0.02 -0.03 -0.01 -0.02  0.00 -0.04  1.00  0.01  0.07
r[8] -0.10  0.02 -0.03  0.00 -0.01  0.00 -0.04  0.01  1.00  0.06
r[9] -0.61  0.10 -0.20 -0.03 -0.09  0.02 -0.25  0.07  0.06  1.00
\end{verbatim}
}
\noindent
The $r_i$ with some sizable correlation with the mass are,
in the order, $r_9$, $r_6$ and $r_3$, i.e. those
related to the individual results that mostly differ from
the barycenter of the final pdf of the mass got by the MCMC sampling
(see Fig.~\ref{fig:hyper_sceptical_a} in order
to get an idea of the reason for the sign of each correlation
coefficient).

\section{Conclusions}
The initial motivation of this paper was didactic, i.e. how
to perform a sceptical combination of results
by MCMC using a convenient program, 
after having got
a better insight of the problem by
{\em Bayesian network} (this is the name also used for the graphical models
we have encountered here).
The choice of the physics case was fortuitous, having been recently
personally interested in the charged kaon mass and having learned
thus about `apparent' disagreements between the most accurate measurements.
However, it is clear that this paper is far from attempting to
give a definite answer,
for which not only a `statistical'\footnote{For example
  it is important 
  to understand how the
  `errors' were evaluated, also because we are aware of the old custom
  (maintained also presently by several experimental teams)
  of using for `systematic errors' extreme variations
  for sake of safety, thus providing very conservative `error',
  instead than standard uncertainties~\cite{ISO,ISOD}. }
but also a serious phenomenological analysis should be required.
For example in Ref.~\cite{Kmass75} there are interesting hints
on not well understood high order
corrections~\cite{Chen,Wilets-Rinker} and it would be
interesting to investigate if the question has been settled down
in the meanwhile and 
what should be the effect on the published mass values, or whether
and how 
its uncertain value should contribute to the overall uncertainty.

The result of this analysis is
$\left.m_{k^\pm}\right|_{I} = 493.677\pm 0.013\,\mbox{MeV}\,,$
  where $I$ stands for all the conditions
  referred in section 6.2 
  ({\em probability is always conditional
  probability} and hence so are also pdf's and moments of distributions).
The result {\em seems} in practical perfect agreement with the
PDG one reminded in Fig.~\ref{fig:PDF2019}. 
But, first, the $f(m\,|\,I)$ estimated by sampling 
is not trivial and definitely far from Gaussian (see
solid thick line of Fig.~\ref{fig:hyper_sceptical_a}), 
yielding e.g. the following {\em probability intervals} (not ``C.L.'s''!):
\begin{eqnarray*}
  P(493.650 \le\, m_{k^\pm}/\mbox{MeV}\, \le 493.697\,|\,I) & = & 95\,\% \\
  P(493.668 \le\, m_{k^\pm}/\mbox{MeV}\, \le 493.687\,|\,I) & = & 50\,\% \\
  P( m_{k^\pm}/\mbox{MeV} \, \le 493.678\,|\,I) & = & 50\,\%\,. 
\end{eqnarray*}
Second, even if the numerical results coincide,
this agreement is just due to a compensation
of two effects in the PDG analysis which go into apposite directions:
\begin{itemize}
\item 
  a weighted average of all nine individual results
  (see Tab.~\ref{tab:masseK_PDG_Gall}),
  with the final `error' scaled according to the
  $\sqrt{\chi^2/\nu}$ prescription, would have lead to
  $493.664\pm 0.011\,$ MeV, that is 13\,keV lower
  than that reported by the PDG~\cite{PDG2019};
\item however, 
  the analysis
  was not performed on the nine individual results of Tab.~\ref{tab:masseK_PDG_Gall},
  but on the six ones of Tab.~\ref{tab:masseK_PDG}, where
  the precise result $493.631\pm 0.007\,$ MeV of Ref.~\cite{Kmass88} 
  had been `weakened' by the other three because of the  $\sqrt{\chi^2/\nu}$
  scaling prescription already applied by the authors.
  For this reason the overall result went up
  to  $493.677\pm 0.013\,$MeV, hence producing a bias of
  $+13\,$keV, that is of the same size of the quote `error'. 
\end{itemize}  
The latter point is the surprising novelty  of
this work, and it deserves another paper~\cite{GdA_combination_no_sufficiency}
and perhaps further investigation to check if other, perhaps more
important results are affected by such a bias too.

\mbox{} \\
It is a pleasure to tank Andrea Messina, Enrico Franco and Paolo Gauzzi
for discussions on the subject and comments on the manuscript.


\newpage
\section*{Appendix -- A case of possibly `too good' agreement}
\begin{figure}[!b]
  \begin{center}
    \epsfig{file=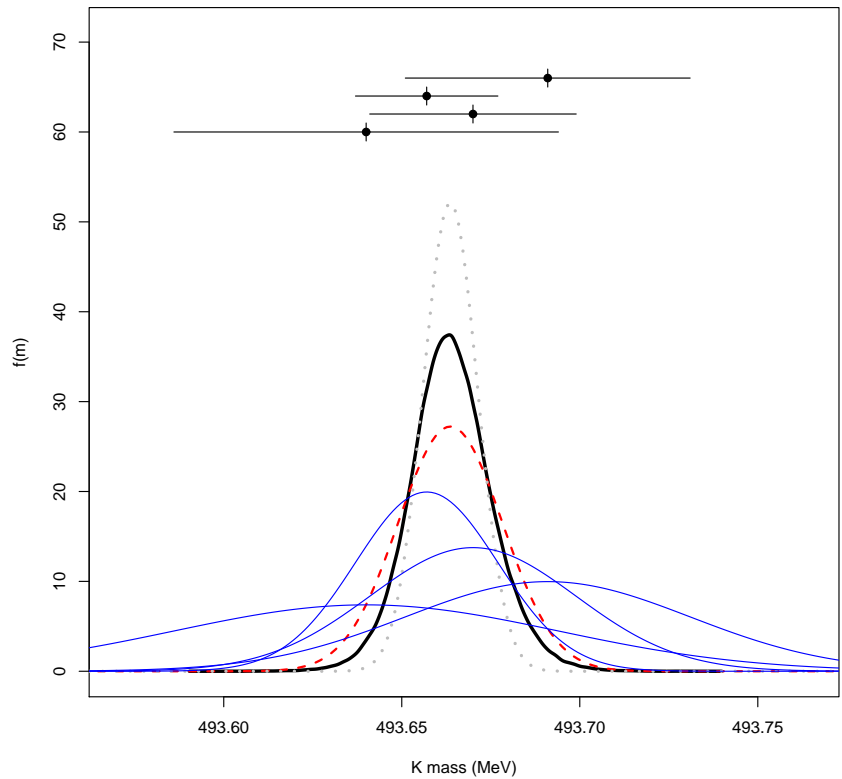,clip=,width=0.73\linewidth}
%
%
%
%
%
  \end{center}
  \caption{\small \sf Standard and sceptical analysis of the results
    published before year 1988.  
  }

\label{fig:sceptical_pre1988}
\end{figure}

Let us also see a case in which the mutual agreement among
individual results `seems' {\em too good}. In order to use again the
kaon mass data, we take the four results published before year 1988,
shown with solid blue Gaussians in
Fig.~\ref{fig:sceptical_pre1988}, to which we over-impose
(usual dashed red Gaussian) the outcome of the
weighted average yielding $493.664 \pm 0.015\,$MeV.
But in this case our suspicion is that the uncertainty could be
overestimated. Indeed, if we calculate the $\chi^2$ we get 0.818,
with a $\chi^2/\nu$ of 0.27 (p-value 0.85). Applying {\em strictly}
the  $\sqrt{\chi^2/\nu}$ scaling prescription -- frequentist {\em gurus}
probably might not agree, but let us go on with the {\em exercise} --
we get a scaling factor of $\times 0.52$, and thus an `error' of
$7.7\,$keV (dotted gray Gaussian). The posterior pdf of the sceptical
analysis (solid thick black line) is this time
practically Gaussian and gives
$493.664\pm 0.012\,$MeV: the curve is narrower than
the simple weighted average,
in agreement with our suspicions, but not as narrow
as when the  $\sqrt{\chi^2/\nu}$ scaling was (improperly?) used.
Conclusions on this last comparisons are left to the reader.

\end{document}